\documentclass[pdflatex,sn-mathphys-num]{sn-jnl}

\usepackage{graphicx}%
\usepackage{multirow}%
\usepackage{amsmath,amssymb,amsfonts}%
\usepackage{amsthm}%
\usepackage{mathrsfs}%
\usepackage[title]{appendix}%
\usepackage{xcolor}%
\usepackage{textcomp}%
\usepackage{manyfoot}%
\usepackage{booktabs}%
\usepackage{algorithm}%
\usepackage{algorithmicx}%
\usepackage{algpseudocode}%
\usepackage{listings}%
\usepackage{subcaption}
\usepackage{caption}
\usepackage[export]{adjustbox}
\usepackage{listings}
\usepackage{hyperref} 

\theoremstyle{thmstyleone}%
\theoremstyle{thmstyletwo}%

\theoremstyle{thmstylethree}%
\newcommand{\E}{\mathbb{E}}

\begin{document}

\title[Article Title]{Network topology effects on the social circle polls}


\author*[1,2]{\fnm{Giovanni} \sur{Palermo}}\email{giovanni.palermo@cref.it}

\author[1,2,3]{\fnm{Emanuele} \sur{Brugnoli}}
\author[1,2,3]{\fnm{Ruggiero D.} \sur{Lo Sardo}}
\author[1,2,3,5]{\fnm{Vittorio} \sur{Loreto}}
\author[2,4,‡]{\fnm{Giulio} \sur{Prevedello}}
\author[2,4,‡]{\fnm{Pietro} \sur{Gravino}}

\affil*[1]{\orgdiv{Physics Department}, \orgname{Sapienza University of Rome}, \orgaddress{\street{Piazzale Aldo Moro 2}, \city{Rome}, \postcode{00185}, \country{Italy}}}

\affil[2]{\orgdiv{Centro Ricerce Enrico Fermi}, \orgaddress{\street{Via Panisperna 89/A}, \city{Rome}, \postcode{00185}, \country{Italy}}}

\affil[3]{\orgdiv{Sony Computer Science Laboratories Rome, Joint Initiative CREF-SONY}, \orgaddress{\street{Centro Ricerche Enrico Fermi, Via Panisperna 89/A}, \city{Rome}, \postcode{00184}, \country{Italy}}}

\affil[4]{\orgdiv{Sony Computer Science Laboratories Paris}, \orgaddress{\street{6, Rue Amyot}, \city{Paris}, \postcode{75005}, \country{France}}}

\affil[5]{\orgdiv{Complexity Science Hub Vienna}, \orgaddress{\street{Metternichgasse 8}, \city{Vienna}, \postcode{1030}, \country{Austria}}}
\affil[‡]{These authors contributed equally to this study}



\abstract{
Election polls play a critical role in political discussions by probing public opinion and enabling political parties to assess their performance before elections. However, traditional polling methods sometimes fail to predict election outcomes accurately, leading researchers to explore new methodologies. One such approach is the ``social circle'' question, which asks respondents about the voting preferences of their social contacts. This method leverages collective intelligence and has shown promise in improving predictive accuracy. Nevertheless, the influence of the social network's topology on the effectiveness of social circle polls remains unexplored.

In this study, we develop a theoretical framework to analyse how social network structure affects polling accuracy. By simulating voter networks with varying levels of polarisation and connectivity, we assess the performance of both standard and social circle polling methods. Our findings indicate that while social circle polls generally outperform traditional approaches, certain network characteristics can introduce biases and undermine their performances, particularly in polarised situations, which are increasingly frequent in the current political landscape. To address these challenges, we propose a new estimator that combines information from both standard and social circle polls, improving election outcome predictions. We demonstrate the applicability of our method using real-world polling data from the 2016 U.S. presidential election, showcasing its practical utility and providing an estimate of the polarisation level in that society.

This work establishes a foundation for enhancing polling methodologies by assessing and integrating network features, which has significant implications for social and political research.
}



\maketitle

\section{Introduction}\label{intro}
The first known example of an opinion poll dates back to 1824, when a tally based on a small sample correctly predicted the election of Andrew Jackson as President of the United States of America~\cite{Tankard1972}. Since the outcome was successful, such straw votes gradually became more popular, but they remained local, usually citywide phenomena.

The first structured polls on a national sample were carried out only in 1916 by the weekly magazine \textit{The Literary Digest}, which managed to predict the outcome of the U.S. Presidential elections five times in a row.~\cite{jstorFirstStraw}.
Later, polls became increasingly popular and became a standard tool of political analysis in the 20th century. They were used by the media to probe the population's leanings and by parties to assess their performance before the elections and during their mandate.
Their reliability has thus important consequences on information and decision-making.
While survey methods have evolved over the years, polls still occasionally fail in predicting the outcome of the elections~\cite{Kennedy2018, Zhou2021}, especially when the difference of votes between the winner and the runner-up is small.
Inaccuracies result from many factors. First, the limited number of respondents that pollsters can interview creates uncertainty in predictions~\cite{Kennedy_Hartig}. Second, the reluctance among individuals to disclose their actual voting intentions~\cite{Bradburn_Sudman_Wansink_2015}, particularly when it involves controversial candidates. Additional factors include the lack of representativeness of survey samples~\cite{battaglia} and the adoption of complex electoral systems, in which election results differ significantly from a simple vote count. For instance, in the so-called ``first-past-the-post'' system used in the UK, where voters elect candidates at the district level, a party with much less than 50\% can still gain the majority of seats in the Parliament, as long as it wins in most districts.

In such cases, guessing the total share of votes for each candidate or party is not enough. A notable occasion in which this resulted in an issue is the election of Donald Trump as President of the USA in 2016. Most polls correctly predicted Clinton would get the majority of votes, and even pollsters working in Trump's campaign confirmed he would have lost the election~\cite{Jacobs}. Nevertheless, Trump was finally elected President because of most states' winner-take-all system. In such cases, being accurate at a local level or with a smaller population is crucial.

To enhance predictive accuracy, pollsters have been formulating new question formats to extract more information from the survey and mitigate sample size limitations.
The ``Election winner'' question is one of these attempts. It asks participants who they think will be elected~\cite{Rothschild2011}, replacing the usual inquiry about their voting intention. In this way, by collecting the persons' perceptions about the vote, the pollsters virtually increase the sample size and exploit a \emph{wisdom of the crowd} effect~\cite{GALTON1907} to predict the winner. While this approach has been successful in certain cases~\cite{LewisBeck1989, Graefe2014}, it also presents some limitations. Notably, it focuses on the election outcome rather than vote percentages, provides results at the national rather than local level, and relies exclusively on public perception, which may be influenced by media and other external factors~\cite{Irwin2002}. In fact, the \emph{wisdom of the crowd} is susceptible to biases and correlations in population responses\cite{DavisStober2014, Lorenz2011}. 

The ``Social circle'' question was introduced to overcome these limitations~\cite{Nisbett1985, Dawtry2015}. An illustrative example of such a question is: ``Of all your social contacts who are likely to vote, what percentage will vote for [party/candidate]?''~\cite{Galesic2018}. This approach offers several advantages: it provides estimates of vote distribution, enables local-level analysis, and is less susceptible to biases introduced by external influences~\cite{Galesic2018}. 
Moreover, this estimator is less affected by individuals' reluctance to disclose their voting intentions, as previous studies have shown that estimates based on social circle align more closely with the actual population than self-reported perceptions~\cite{Galesic2012}.

Initial trials of political surveys indicate that social circle polling may surpass traditional polling in accuracy~\cite{Galesic2018}, even in multi-party systems~\cite{BruinedeBruin2022}. However, for the methodology to become a new standard, it requires validation across a broader range of contexts and support by a robust theoretical framework.
The existing literature has focused on studying the network of voters at the scale of the individual participant connections~\cite{Lee2019, AhlstromVij2022}.
We argue that, since this methodology leverages social connections, understanding the impact of the entire network structure is of paramount importance.

In this work, we evaluate social circle polling on a network of voters, examining the impact of topological features on performance and identifying ways to improve estimates. We introduce a theoretical framework that employs a stochastic model to generate networks with varying structural properties, enabling simulations of both classic and social circle polls under different configurations. By calculating the risk and accuracy of these estimators, we measure how much topological polarisation can decrease the performance of social circle polls compared to standard polls.

By leveraging our theoretical framework, we could estimate the network's polarisation with high accuracy. Using this new information, we developed an improved estimator that outperforms both classic and social circle poll estimators.
Finally, we demonstrate our methodology on two surveys from the 2016 US presidential elections, finding statistical compatibility between their two estimates of polarisation. Our work lays the groundwork for leveraging social circle surveys to improve the accuracy of public opinion assessment while uncovering structural properties of society.

\section{Results}\label{results}
\subsection{Theoretical framework}\label{subsec:teoretical_framework}
To quantify the performance of social circle-based polls over the standard polls, we start by presenting the mathematical framework to simulate a population and formalize these surveys. 

\subsubsection{The network of voting population}
\begin{figure}[h]
    \begin{subfigure}[t]{0.32\textwidth}
        \centering
        \includegraphics[width=\linewidth, valign=T]{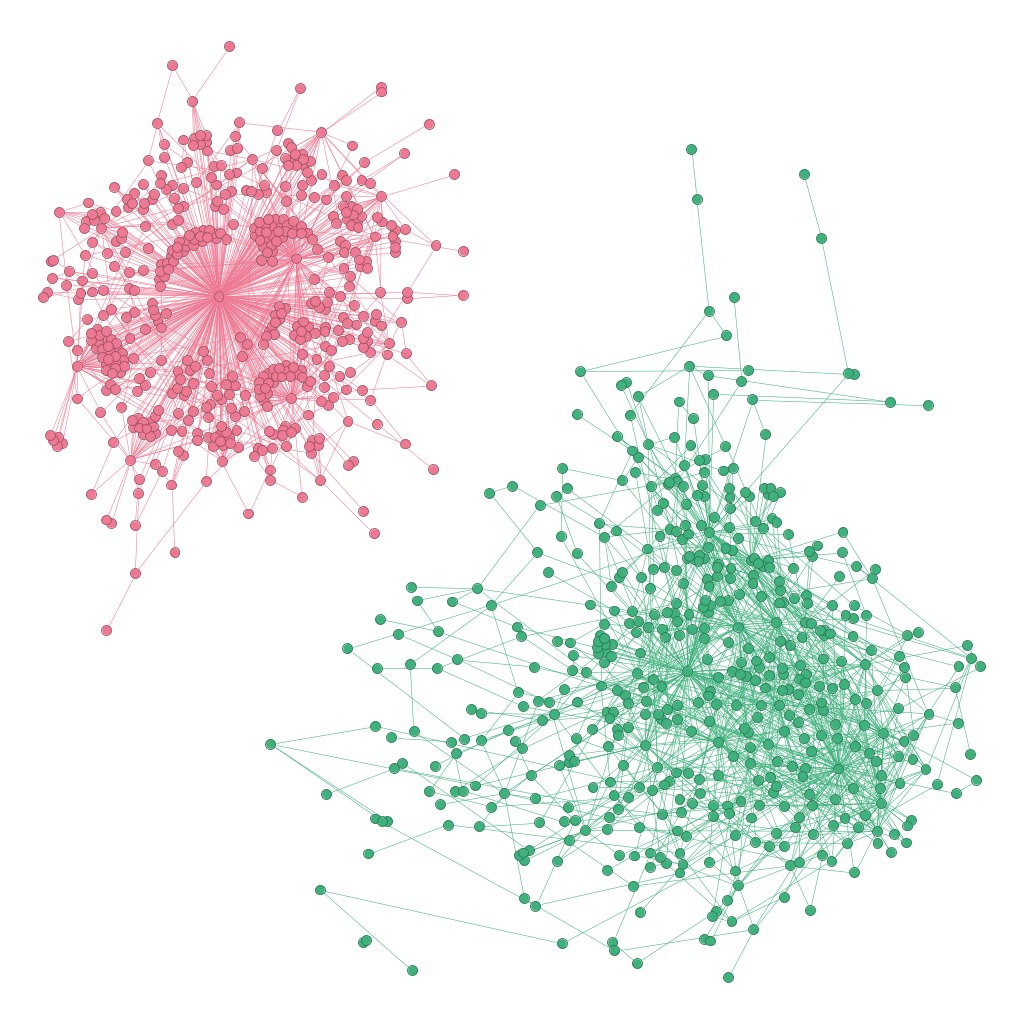}
        \subcaption{$q=0$}
        \label{graph1}
    \end{subfigure}    \begin{subfigure}[t]{0.32\textwidth}
        \centering
        \includegraphics[width=\linewidth, valign=T]{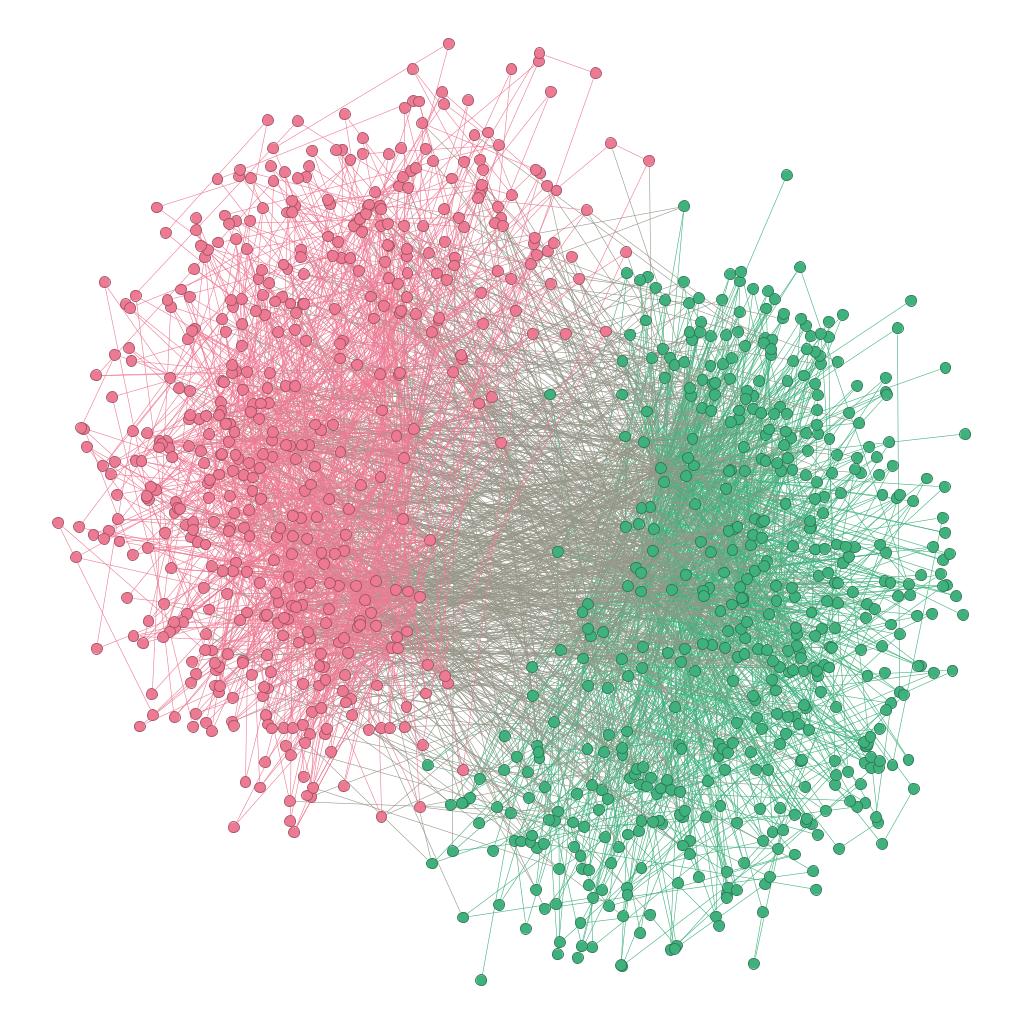}
        \subcaption{$q=0.25$}
        \label{graph2}
    \end{subfigure}    \begin{subfigure}[t]{0.32\textwidth}
        \centering
        \includegraphics[width=\linewidth, valign=T]{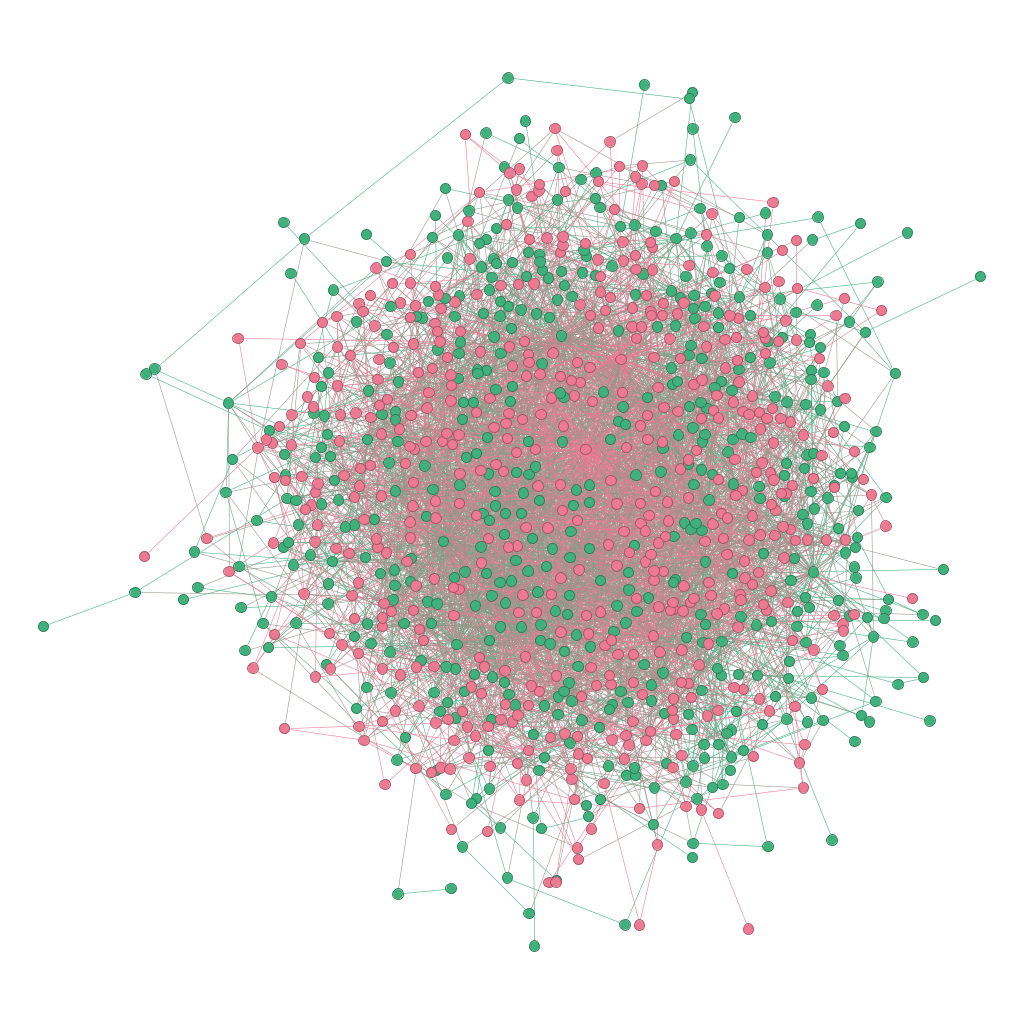}
        \subcaption{$q=0.5$}
        \label{graph3}
    \end{subfigure}
    \caption{Example graphs generated using the Stochastic Block Model (SBM) with $N=1000$, $M=0$ and $\kappa=4$. Nodes are coloured green and red to represent vote intentions of $+1$ and $-1$, respectively. The parameter $q$ controls the network topology. At $q=0$ (a), the communities are completely disconnected. As $q$ increases (b), more connections are established between the communities. At $q=0.5$ (c), the communities merge into a single block. For $q>0.5$, networks become increasingly heterophilic, culminating in a bipartite structure at $q=1$ where each node connects exclusively to nodes in the other community.}
    \label{network}
\end{figure}

We focus our analysis on a population structured as a network of acquaintances who can vote for one of two options.  
This simple setup covers many real situations, like a referendum with two voting options.
We formalise our social network as an undirected graph $G = (N, E)$ where $N$ is the size of the population and $E$ is the number of connections between two individuals.
In our graph, each person $i$ is randomly connected to $k_i\geq0$ neighbours that constitute the person's social circle. 
We denote the average degree of the network as $\kappa=2E/N$, indicating the density of connections in the graph. 
The $i$-th person's vote is denoted with $x_i\in \{-1, 1\}$, dividing the population into those who prefer one option or the other.
Then, the result of the election is determined by the average vote
\begin{equation}
    M=\frac{1}{N}\sum_{i=1}^{N}x_i,
\end{equation}
that is also the magnetisation of the system that we want to estimate.
For example, $M=\pm1$ means the totality of the network votes $\pm1$, while $M=0$ is a perfect tie.
The sizes of the communities of positive and negative voters, $N_+$ and $N_-$ respectively, can then be calculated as $N_\pm = \frac{1\pm M}{2}N$.
Key to our analysis is examining the number of links between nodes with different opinions, quantified by $E_C$, and the ratio $q=E_C/E$, representing the proportion of links connecting voters with opposing views.
The ratio $q$ can be interpreted as a measure of polarisation. If $q=0$, the communities are completely disconnected, resulting in a totally polarised situation. In contrast, when $q=1$, all links connect opposing groups, depicting a totally heterophilic network (see Fig.~\ref{network}) such as a bipartite graph.
To generate graphs for our analyses, we adopted the Stochastic Block Model (SBM), as it meets these requirements and is well-known in the literature for generating networks with adjustable degrees of polarisation~\cite{Lee2019_sbm}. Such a model depends on parameters $N$, $M$, $\kappa$, and $q$ (see Methods), allowing us to study their impact on the estimators' performances.

\subsubsection{Polling Methods}
To simulate polling of the population, we randomly sample individuals with uniform probability to create a sample set $Z$ of size $n<N$.
While real-world polling typically involves a more refined sampling process aimed at ensuring representativeness with respect to demographic factors such as age, gender, geographic location, and other characteristics influencing voting intentions, our simulations focus on the simplified case of uniform random sampling. This approach allows us to isolate and examine the topological effects on the social network. Importantly, the results of this analysis are extendable to scenarios that more closely align with sampling techniques employed by pollsters, as will be shown later.
For each randomly sampled node $i$, we collect its voting intention $x_i$ and the average voting intention of its neighbours, $s_i$, computed as
\begin{equation}\label{eq_sc}
    s_i=\sum_{j=1}^{N}\frac{A_{ij}x_j}{k_i}
\end{equation}
where $A$ is the adjacency matrix of the graph $G$. This equation assumes each voter can accurately report the average opinion of their neighbourhood. This assumption neglects any bias the person can have in reporting his social circle opinion.  While there is some evidence people might be affected by cognitive biases~\cite{Goel2010}, this choice allows us to simply model the social circle and isolate how the network topology affects its performance.

Given the sample $Z$, we can predict the magnetization of the system in two ways. As in classic polls, we can calculate the average of the voting intentions
\begin{equation}
    \overline{x}=\frac{1}{n} \sum_{i \in Z}x_i,
    \label{eq:est_cl}
\end{equation}
The other estimation is calculated by computing the average social circle voting intention
\begin{equation}
    \overline{s}=\frac{1}{n} \sum_{i \in Z}s_i,
    \label{eq:est_sc}
\end{equation}
where overline notation, hereafter, denotes sample averaging.
These two estimators form the basis of our analysis. Specifically, we aim to evaluate their effectiveness in predicting the magnetization $M$ of the network as a function of network parameters.
While $\overline{x}$ is unbiased and its distribution only depends on $N_\pm$ and $N$, converging to $M$ as $N$ increases, we anticipate the social circle prediction $\overline{s}$ is biased and also affected by the network topology, as it depends on the connections between voters and the connections' stochasticity. 
It is worth noting the parameter $n$ only concerns the polling phase.

\subsection{The parameters space}
We evaluated the performance of $\overline{x}$ and $\overline{s}$ on networks generated via SBM under various parameter configurations. In particular, we explored the scenarios for: $M \in [-0.1,0.1]$, corresponding to a maximum difference of $10\%$ in the voting results, a range in which polling accuracy becomes critical; $q \in [0.1,0.5]$, to account for a reasonable degree of average polarisation, as almost disconnected ($q<0.1$) or heterophilic communities ($q>0.5$) would be rather unrealistic.
We set $N=10^5$ (the size of a large city), $n=10^3$ (a sample of $1\%$ of the population) and $\kappa=50$, corresponding to the average number of an individual’s supportive group as theorised by Dunbar~\cite{Dunbar1992}. Varying these values did not substantially affect the results (see SI).

\begin{figure}[H]
   \centering
   \includegraphics[width=\linewidth]{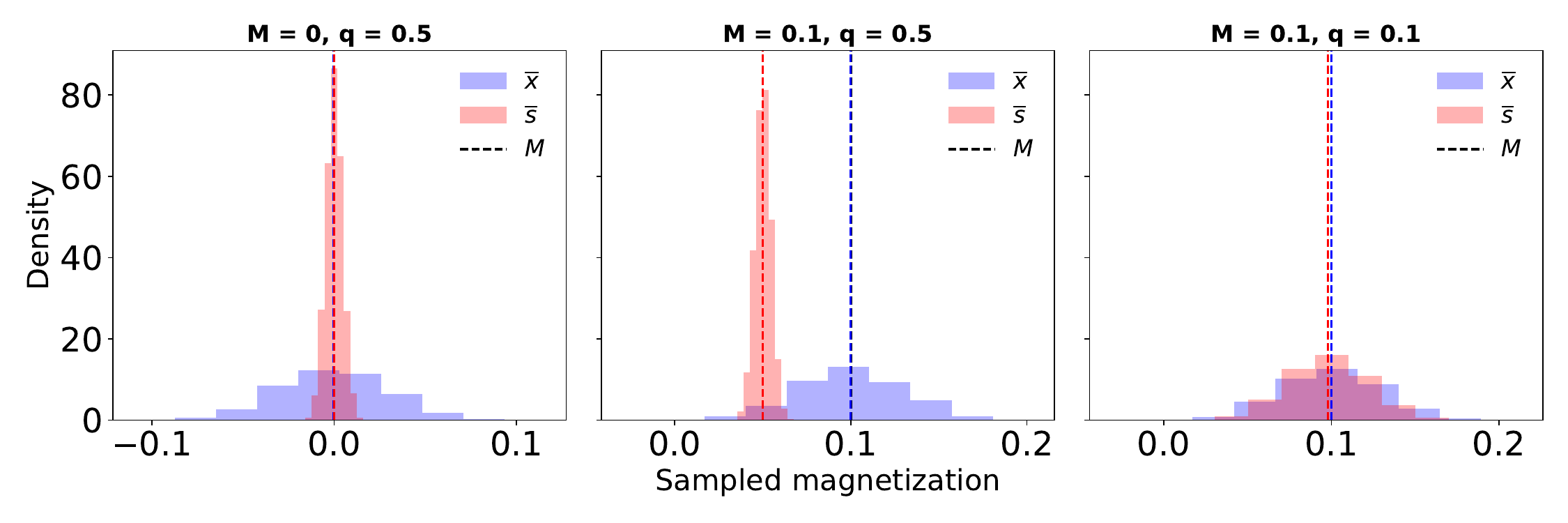}
   \caption{The distributions of standard ($\overline{x}$, blue) and social circle ($\overline{s}$, red) polling methods from $10^4$ simulations on a network of size $N=10^5$ with $\kappa=50$ for different values of $M$ and $q$. The sample size was fixed at $n=10^3$. Dashed vertical lines indicate the actual magnetisation ($M$, black) and the average of the statistics over all simulations for both polling methods. The plot shows how the mean and variance of the social circle statistic, compared to the normal poll, are influenced by variations in magnetisation and topology.
   }
   \label{multi_hist}
\end{figure}

For each parameter configuration, we generated $10^4$ synthetic networks, each surveyed via standard and social circle methodology to infer the distributions of $\overline{x}$ and $\overline{s}$. In Fig.~\ref{multi_hist}, we show the distributions for three example configurations. It is evident how bias and variance of social circle polling show different behaviour when compared to standard polling.

\subsection{Assessment of the Estimators}
\begin{figure}[!http]
        \centering
        \includegraphics[width=\linewidth, valign=T]{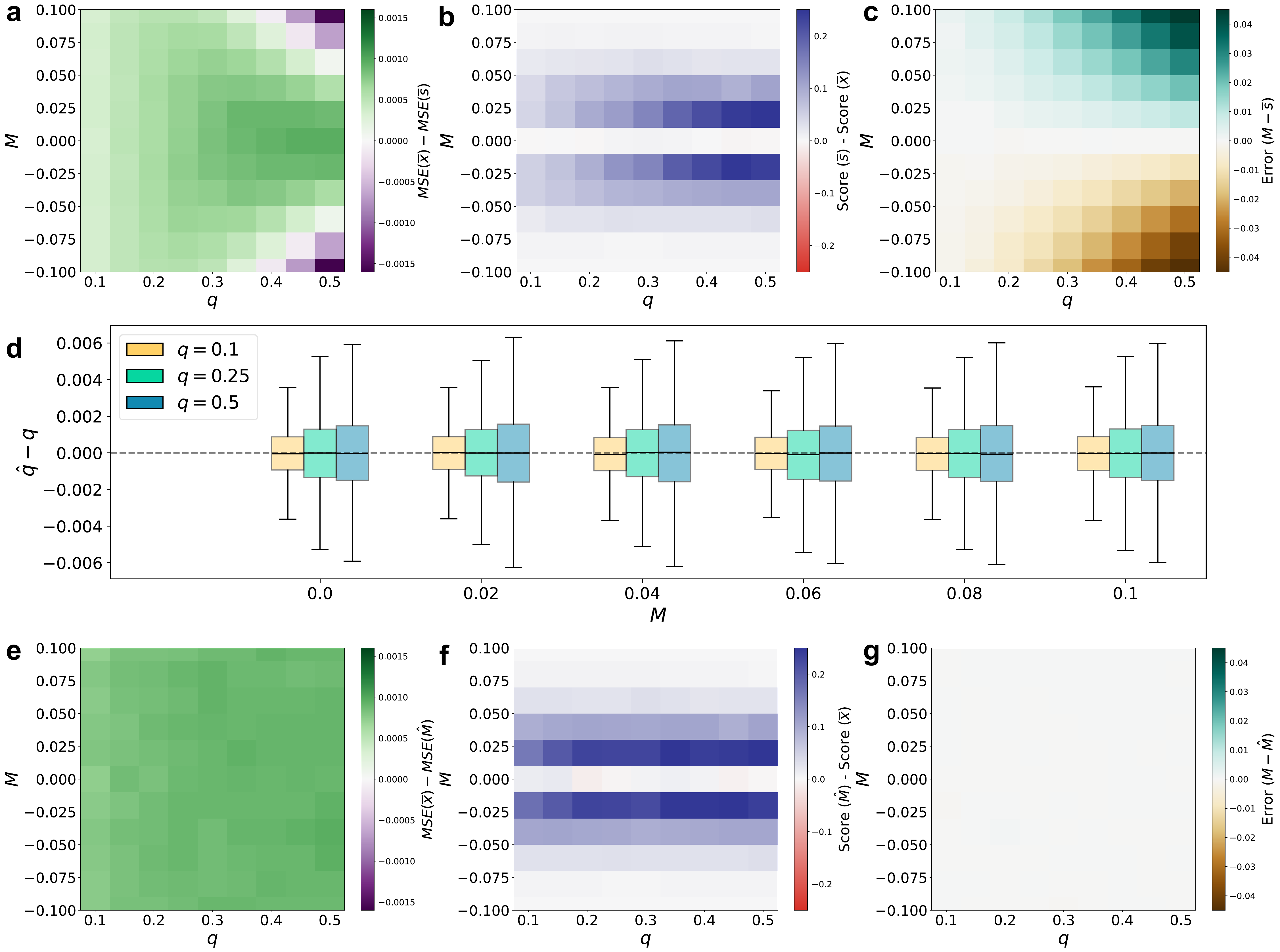}
        \label{figure3}

    \caption{Estimators performance. (a) The difference in Mean Squared Error (MSE) between standard and social circle estimators. Areas where the social circle estimator is less risky than the standard are green; otherwise, they are violet.
    (b) Difference in accuracy between standard and social circle estimators. Areas where the social circle estimator more frequently predicts the winner are blue; otherwise they are red.
    (c) Bias of the social circle statistic in estimating the real magnetization.
    (d) Boxplots for the distribution of the error of statistic $\hat{q}$ in estimating $q$. Full lines indicate the medians, while the box represents the interquartile range, and whiskers extend from the 0th to the 4th quartile (excluding outliers).
    Despite a slight increase in the error as $q$ grows, the errors remain within $\pm 0.01$ most of the times.
    Panels (e-g) correspond to panels (a-c), but with the adjusted social circle statistic $\hat{M}$ replacing the former social circle statistic. 
    All plots are based on network simulations with parameters $N=10^5$, $\kappa=50$ and $n=10^3$.}
    \label{fig:heatmap}
\end{figure}

To systematically compare the performance of the estimators $\overline{x}$ and $\overline{s}$, we calculated three assessment metrics based on samples from networks simulated via SBM.
First, we measured the difference in Mean Squared Error (MSE) between the two estimators to quantify their relative risk, accounting for both bias and variance (Fig.~\ref{fig:heatmap}a). 
Second, we calculated the difference in Rand Index (RI) between the estimated magnetisation, assessing which estimator more frequently identifies the election winner, irrespective of the margin of victory (Fig.~\ref{fig:heatmap}b).
Third, we examined the estimation error of $\overline{s}$ to determine the parameter regimes in which its bias is most pronounced (Fig.~\ref{fig:heatmap}c).

We observed that social circle polling outperforms standard polling in most cases.
Specifically, the social circle estimator becomes less risky as $q$ increases and $m$ decreases simultaneously. However, when $q>0.4$, the social circle statistic exhibits a higher risk for larger absolute values of $M$. Despite this, the social circle estimator is always better or comparable to the standard one in terms of predicting the winner, with improvements reaching up to $20\%$.

These results indicate that, at least in the presence of two similar communities, social circle polling is preferable to standard polling, with performance gains that vary based on the degree of polarisation. In fact, we observe that as polarisation rises, i.e. $q$ decreases, the performance of the social circle estimator progressively declines towards the level of the standard estimator.
Thus, to assess these gains, it is essential to determine $q$, which requires information about network connections, typically unavailable in real-world survey settings.

\subsection{Estimating topological polarisation}\label{subsec:heterophily}
Given its crucial role in the performance analysis, we tackled the estimation of $q$, which represents the fraction of links connecting nodes with opposing voting intentions. This value measures the network's topological structure and quantifies the degree of polarisation within the population. We demonstrated that $q$ can be estimated as the harmonic mean of $q_+$ and $q_-$, where $q_\pm$ are the fractions of stubs from links connecting nodes from each community to the other (see Methods). To estimate $q_\pm$ we will leverage the concept of heterophily. 

For an individual $i$, we define heterophily as the proportion of $i$'s neighbours, in the social network, who hold opposing voting intentions, which can be calculated as follows:
\begin{equation}
\eta_i=\lvert \frac{x_i-s_i}{2} \rvert \in [0,1].
\end{equation}
Averaging heterophily over each community provides the mean probability that a randomly selected stub from a node in one community connects to a node in the other community.

Averaging individual heterophily within a community provides the proportion of stubs from a randomly selected node in that community that connects to the other community, which corresponds to $q_\pm$.
Thus, by estimating $q_+$ and $q_-$ with the average heterophily within the subsamples $x_i = +1$ and $x_j = -1$,i.e. $\overline{\eta}_{\pm}$ respectively, we define the estimator for $q$ as:
\begin{equation}
    \hat{q}=\frac{2}{\frac{1}{\overline{\eta}_+}+\frac{1}{\overline{\eta}_-}}
    \label{eq_q_est}
\end{equation}

We evaluated the performance of $\hat{q}$ using the simulations of the polling problem under study (Fig.~\ref{fig:heatmap} d). This analysis showed that the estimate of $q$ has a small error. Thus $\hat{q}$ is sufficiently accurate to identify the risk profile from the heatmaps of Fig.~\ref{fig:heatmap}. Furthermore, the estimate of $q$ provides a direct measure of topological polarisation, that is the level of segregation between the two communities in the network. Beyond its intrinsic value in characterizing the graph's structure, this measure can also enhance the accuracy of our estimation of $M$, as shown in the following section.

\subsection{Adjusted social circle}\label{correction}

Given that the estimator $\hat{q}$ provides valuable insight into the relationship between the standard statistic and the social circle statistic, we sought to investigate whether combining this information with the polling methods results ($\overline{x}$ and $\overline{s}$) could improve the estimation of network magnetization $M$, particularly under the assumption of a network generated by the SBM.

The approach is to condition the social circle estimator on $\overline{x}$, quantifying the extent to which the surveyed sample divides between voting options. The intuition is that the value of $\overline{s}$ correlates with $\overline{x}$, and shifts accordingly.
Then $\hat{q}$ was used to measure network polarisation, so to infer — and potentially mitigate — the bias in $\overline{s}$.

Based on this framework, we derived an adjusted version of the social circle estimator (see Methods):
\begin{equation}
    \hat{M}=\frac{\overline{s}-\overline{x}(1-2\hat{q})}{2\hat{q}(1-\hat{q})}.
    \label{eq_corr}
\end{equation}

Performance analysis showed that $\hat{M}$ consistently outperforms standard polling across all tested cases (Fig.~\ref{fig:heatmap} e-g), providing a viable alternative to the standard estimator that is not affected by variations in polarisation ($q$).
Further analysis indicated that $\hat{M}$ behaviour is weakly influenced by parameters $\kappa$ and $n$ (see SI). The only case in which its performances dramatically drop is when the average degree of the network is low ($\kappa \lesssim 10$), considerably far from Dunbar's observations.




\subsection{Testing on Real Data}\label{real data}
To test our insights on social circle polling and demonstrate our methodology for estimator evaluation, we considered two pre-election surveys from GfK and USC Dornsife, previously analysed in~\citet{Galesic2018}. These surveys sampled the American population in the week before the 2016 U.S. presidential election, using both questions of standard polling and social circle polling. To align the data with our two-choice framework, we excluded third-party preferences, focusing only on respondents supporting Donald Trump, $x_i=+1$, or Hillary Clinton, $x_i=-1$ (see Methods). In the actual election, the popular vote yielded a network magnetization of $M=-0.022$.

For each survey, we calculated estimates, MSEs and MSDs (Mean Squared Deviation) from the election results ($M$) for the standard $\overline{x}$, social circle $\overline{s}$, and adjusted social circle $\hat{M}$, and topological polarisation $\hat{q}$ statistics (Table~\ref{tab:polls}). To improve the representativeness of the samples of the American population, the means used in the statistical calculations were replaced with weighted means, where the weights were derived from demographic data and provided in the surveys (see Methods). These weights were also used in the estimate of Confidence Intervals (CIs).

\begin{figure}[H]
    \centering
    \includegraphics[width=\linewidth]{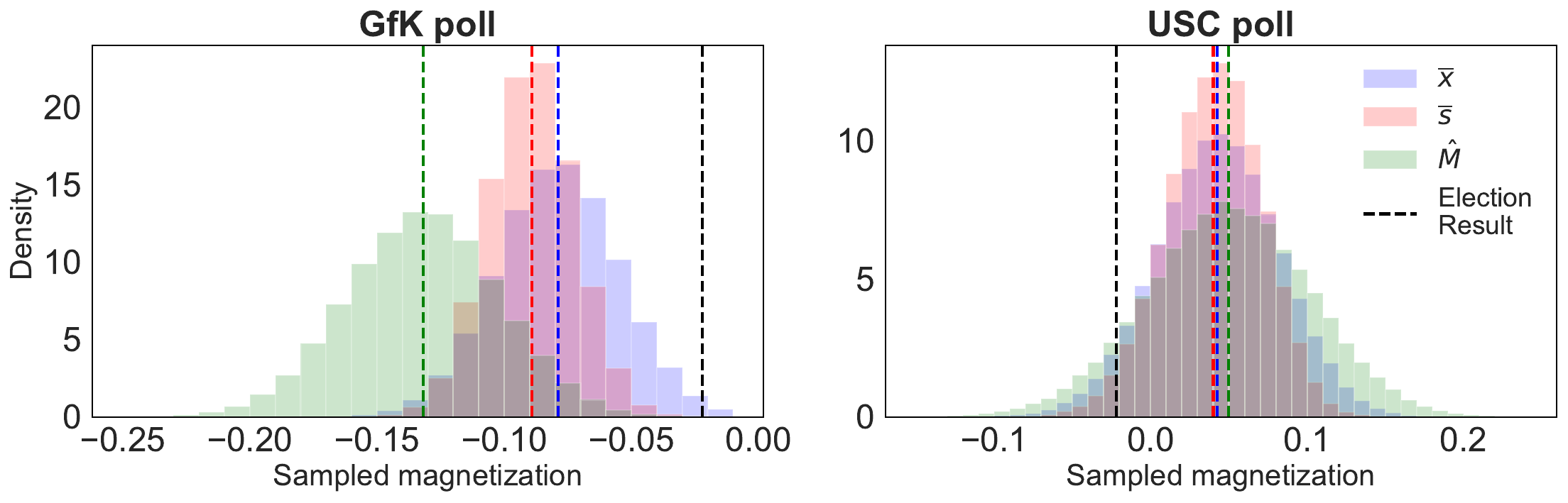}
    \caption{Bootstrap distributions from the GfK (left) and the USC (right) polls for the statistics: $\overline{x}$ (blue), $\overline{s}$ (red) and $\hat{M}$ (green). Dashed vertical lines indicate the actual election results (black) and the statistics observed from the data.}
    \label{fig:usc_gfk}
\end{figure}

By inspecting the MSD from $M=-0.022$, $\overline{x}$ provided the closest estimate to the election outcome in the GfK survey, whereas $\overline{s}$ was the closest for the USC survey. The adjusted social circle statistic, $\hat{M}$, appeared to overcorrect the bias present in the social circle statistic, shifting the prediction away from the electoral results (Fig.\ref{fig:usc_gfk}). The estimates of $q$ were $0.233$ and $0.245$, respectively. These values fall within the range where, based on the risk analysis shown in Fig.\ref{fig:heatmap}, social circle-based estimators outperform the standard polling method in terms of MSE. While this is true for $\overline{s}$ on both surveys, it does not hold for $\hat{M}$.
It is important to note that this analysis is based on only two experiments, unlike the $10^4$ simulations conducted previously. Therefore the individual results of these two experimental instances cannot prove or disprove the correctness of our framework, but only demonstrate its application.

To further investigate the differences between the two polls, we conducted a statistical analysis to test whether the mean statistics were consistent with the hypothesis that they originated from the same population (Table~\ref{tab:polls_hptest}, see Methods). The analysis showed a statistically significant difference in both $\overline{x}$ and $\overline{s}$ between the two surveys, though with a small effect size. This finding aligns with the mixed results observed in the performance analysis discussed earlier. Testing the same hypothesis for $\overline{\eta}_+$ and $\overline{\eta}_-$ showed that differences in heterophily among those declaring the same vote intention were not statistically significant. This result implies that also the $q$ estimates from the two polls are compatible since $\hat{q}$ is a function of solely $\overline{\eta}_+$ and $\overline{\eta}_-$. Consequently, by combining information on individuals' own vote intentions with those from their social circles, we could infer the polarisation of American society shortly before the 2016 election: on average, an individual's social network consisted of approximately $24\%$ of acquaintances with opposing voting preferences (i.e., $q \approx 0.24$), indicating a moderate yet non-negligible level of political diversity within social circles.

\begin{table}[!ht]
    \centering
    \captionsetup{width=\linewidth} 
    \begin{tabular}{lrrrrrrrr}
    \hline
    \multicolumn{1}{|l|}{}              & \multicolumn{4}{c|}{\textbf{GfK}}                     & \multicolumn{4}{c|}{\textbf{USC}}                     \\ \hline
    \multicolumn{1}{|l|}{Est.}              & \multicolumn{1}{r}{Value}                     & \multicolumn{1}{r}{$95\%$ CI}           & \multicolumn{1}{r}{MSE}   & \multicolumn{1}{r|}{MSD}
    & \multicolumn{1}{r}{Value}                     & \multicolumn{1}{r}{$95\%$ CI}           & \multicolumn{1}{r}{MSE}    & \multicolumn{1}{r|}{MSD} \\ \hline
    \multicolumn{1}{|l|}{$\hat{q}$}
        & \multicolumn{1}{r}{0.233} & \multicolumn{1}{r}{(0.222, 0.243)} & \multicolumn{1}{r}{$< 10^{-4}$}  & \multicolumn{1}{r|}{NA}
        & \multicolumn{1}{r}{0.245} & \multicolumn{1}{r}{(0.225, 0.263)} & \multicolumn{1}{r}{$< 10^{-4}$}  & \multicolumn{1}{r|}{NA}  \\ \hline
    \multicolumn{1}{|l|}{$\overline{x}$}
        & \multicolumn{1}{r}{-0.079} & \multicolumn{1}{r}{(-0.126, -0.032)} & \multicolumn{1}{r}{0.0006} & \multicolumn{1}{r|}{0.0038}
        & \multicolumn{1}{r}{0.042} & \multicolumn{1}{r}{(-0.033, 0.121)} & \multicolumn{1}{r}{0.0015} & \multicolumn{1}{r|}{0.0057}   \\ \hline
    \multicolumn{1}{|l|}{$\overline{s}$}
        & \multicolumn{1}{r}{-0.089} & \multicolumn{1}{r}{(-0.122, -0.056)} & \multicolumn{1}{r}{0.0003}  & \multicolumn{1}{r|}{0.0048}
        & \multicolumn{1}{r}{0.040} & \multicolumn{1}{r}{(-0.017, 0.103)} & \multicolumn{1}{r}{0.0009}  & \multicolumn{1}{r|}{0.0048}   \\ \hline
    \multicolumn{1}{|l|}{$\hat{M}$}
        & \multicolumn{1}{r}{-0.132} & \multicolumn{1}{r}{(-0.190, -0.073)} & \multicolumn{1}{r}{0.0009} &  \multicolumn{1}{r|}{0.0129}
        & \multicolumn{1}{r}{0.050} & \multicolumn{1}{r}{(-0.051, 0.154)} & \multicolumn{1}{r}{0.0027}  & \multicolumn{1}{r|}{0.0079}   \\ \hline
    \end{tabular}
    \caption{Estimates, $95\%$ Confidence Intervals, MSE and MSD for the statistics estimating $q$ and $M$ calculated from the GfK and USC survey data (vertical dashed lines). CIs, MSE and MSD from the election outcome ($-0.022$, vertical black line) were calculated via bootstrap procedure (see Methods). MSD for $\hat{q}$ was unavailable due to the absence of a reference value.}
    \label{tab:polls}
\end{table}

\begin{table}[!ht]
    \centering
        \captionsetup{width=\linewidth} 
\begin{tabular}{lrrrrr}
\hline
\multicolumn{1}{|l|}{Estimator}
    & \multicolumn{1}{c|}{$\overline{x}$}
    & \multicolumn{1}{c|}{$\overline{s}$}
    & \multicolumn{1}{c|}{$\overline{\eta}_+$}
    & \multicolumn{1}{c|}{$\overline{\eta}_-$}
    \\ \hline
\multicolumn{1}{|l|}{Statistic}
    & \multicolumn{1}{r|}{-0.086}
    & \multicolumn{1}{r|}{-0.132}
    & \multicolumn{1}{r|}{0.007}
    & \multicolumn{1}{r|}{-0.071}
    \\ \hline
\multicolumn{1}{|l|}{p-value}
    & \multicolumn{1}{r|}{0.007}
    & \multicolumn{1}{r|}{$<$0.001}
    & \multicolumn{1}{r|}{0.860}
    & \multicolumn{1}{r|}{0.199}
    \\ \hline
\multicolumn{1}{|l|}{Effect size}
    & \multicolumn{1}{r|}{-0.001}
    & \multicolumn{1}{r|}{-0.002}
    & \multicolumn{1}{r|}{0.0002}
    & \multicolumn{1}{r|}{-0.002}
\\ \hline                         
\end{tabular}
    \caption{Statistical comparison of estimators between GfK and USC polls by bootstrapping Welch's $t$-test statistic. Statistical significance is measured by the test's two-side $p$-value and the effect size is measured by Cohen's $d$.}
    \label{tab:polls_hptest}
\end{table}

\section{Discussion}\label{discussion}

In this study, we investigated how social circle polling is influenced by the topology of the surveyed population's social network. This revealed new opportunities for predicting election outcomes and measuring polarisation in society.

To study network effects, we developed a theoretical framework for a binary voting scenario and generated stochastic networks of voters under various configurations, with a particular focus on the level of polarisation.

By comparing social circle polling with standard polling in terms of risk, accuracy, and bias, we found that the performance gain of social circle polling depends on both network polarisation and the relative sizes of the two communities. Specifically, as polarisation or community size unbalance increased, the mean squared error (MSE) of social circle estimates also rose, eventually matching or even exceeding the MSE of standard polling estimates. However, social circle polling proved particularly advantageous in scenarios of high uncertainty, where votes were evenly split, as its bias was offset by a lower variance, resulting in superior performance over standard polling.

Leveraging the clear relationship between social circle polling and respondents' heterophily (defined as the proportion of respondent's contacts who vote differently than that respondent), we developed an estimator of topological polarisation that is both accurate and precise. This estimator serves three crucial functions. First, it provides a direct estimate of population polarisation, a key metric in various fields, including network science, sociology, and political science. Second, it enables a systematic risk assessment of social circle estimate over standard polling. Third, building on our theoretical framework, this polarisation estimator enables the integration of standard and social circle polling, yielding an improved estimator of election outcomes with enhanced performance across all simulated scenarios.

Finally, we demonstrated our findings on two surveys from the 2016 U.S. presidential election. Through a bootstrap procedure, we found that the social circle estimator outperformed the standard approach in both cases in terms of MSE. Yet, this was not the case for the improved estimator. When comparing the estimators against the actual election outcome, the results were mixed: social circle polling performed best in one survey, while standard polling was superior in the other. 


Comparing the two surveys using bootstrap hypothesis testing revealed that both the standard and social circle estimators differed statistically, though with a small effect size, while the heterophily estimates for each side remained statistically compatible. This suggests that, despite differences in vote distribution between the two surveyed samples, the structures of social connections across opposing factions were comparable. We speculate that this occurs because opinions fluctuate more among respondents with higher heterophily. In such a scenario, opinion shifts would have a smaller impact on average heterophily than on average opinion. This aligns with the plausible hypothesis that individuals with lower homophily are more susceptible to opinion fluctuations.

Furthermore, using our polarisation estimator, we estimate that the average American had approximately $24\%$ of acquaintances with opposing voting preferences shortly before the 2016 election, indicating a low level of political diversity within social circles.


These results have the potential to improve election outcome forecasts and open new avenues for investigating polarisation across different contexts, including varying time periods and countries. 
Still, while social circle polling consistently outperformed standard polling in most simulated trials, the improved estimator did not perform well on the survey data used to demonstrate the methodology, inviting a reflection on the proposed theoretical framework and its limitations. First, we assumed that individuals accurately report their neighbours' voting intentions when answering the social circle question. This assumption is simplistic, as it does not account for stochastic variations or cognitive biases that may influence responses~\cite{Goel2010, Simon1955, Tversky1974, BOCCALETTI2006, Cohen2002}.
Additionally, we considered only the specific case where the number of contacts an individual has follows an exponential distribution and both opinion groups have the same average number of internal contacts within their respective groups. In real-world settings, however, these assumptions may not hold ~\cite{Cossard2020, Newman2005}. Nonetheless, given the lack of additional information about the topology of social networks of the surveyed population, we argue that this assumption serves as a practical and reasonable starting point.
Future research could focus on identifying the most appropriate distribution for the number of social contacts and deriving the best estimates –directly from polling data– for the average in-group and out-group connections within each community. This would help validate initial assumptions and further enhance the accuracy of network-based polling methods.

By exploring the link between social networks and polling strategies, our findings provide insights into how people's perceptions, filtered through their social connections, can be leveraged to draw a more accurate representation of public opinions while also deepening our understanding of the social network itself. Beyond improving election predictions, this research has broader implications for understanding social influence, polarisation, and information spreading. As societies become more interconnected, and more polarised, developing better methods to measure public opinion will be essential, not just for politics but for tackling wider societal challenges.

\section{Methods}\label{methods}
\subsection{Graph generation}\label{method:graph_generation}
For our study, two-community graphs were generated using the Stochastic Block Model (SBM)~\cite{Holland1983}, as implemented in the Python library \texttt{igraph}, which allows simultaneous control over the number of nodes $N$, magnetisation $M$, average degree $\kappa$, and polarisation $q$.
The used algorithm generates graphs with specified community sizes and connection probabilities between the nodes of the communities. 
The size of the two communities, $N_{\pm}$, were computed from the magnetization and the total number of nodes: $N_{\pm}=\frac{1 \pm M}{2}N$.

The probabilities of connection between and within the two communities are represented by a probability matrix
\begin{equation}
    P=\begin{pmatrix}
        p_{+} & p_{+-} \\
        p_{-+} & p_{-}
    \end{pmatrix},
\end{equation}
where, for undirected graphs as in the present work, $p_{+-}=p_{-+}=p_C$.
With this notation, given two nodes $i$ and $j$ with opinions $x_i$ and $x_j$, the edge between them exists with probability $p_+$ if $x_i = x_j = +1$, $p_-$ if $x_i = x_j = -1$, or $p_C$ if $i$ and $j$ do not belong to the same community.
In this work, we assumed that the average internal degree of the nodes of the two communities is identical, i.e. $\kappa_{+}=\kappa_{-}$. Consequently, it can be expressed in terms of parameters $\kappa$ and $q$ as:
\begin{equation}
    \kappa_{\pm} = \frac{2(E-E_C)}{N} = \kappa(1-q).
\end{equation}
The number of internal edges within each community is then given by:
\begin{equation}\label{eq:epm}
    E_{\pm} = \frac{1}{2}\kappa_{\pm} N_\pm = \frac{1}{2} \kappa(1-q) N_\pm.
\end{equation}
The probability of an internal connection within a community is defined as the number of internal edges divided by the number of possible internal links, that is:
\begin{equation}
    p_{\pm} = \frac{\frac{1}{2}\kappa(1-q) N_\pm}{\frac{1}{2}N_\pm(N_\pm-1)} =  \frac{\kappa(1-q)}{N_\pm-1} \approx \frac{\kappa(1-q)}{N_\pm},
\end{equation}
where the last approximation holds for $N_\pm \gg 1$, namely in all practical cases.
Similarly, the probability of a cross-community connection is defined as:
\begin{equation}\label{eq:ec_pc}
    p_{C} = \frac{\frac{1}{2} q \kappa N}{N_+N_-}
\end{equation}

Finally, the probability matrix can be expressed as a function of the initial parameters as follows:
\begin{equation}
    P=\begin{pmatrix}
        \kappa\frac{1-q}{N_+} & \kappa\frac{qN}{2N_+N_-} \\
        \kappa\frac{qN}{2N_+N_-} & \kappa\frac{(1-q)}{N_-}
    \end{pmatrix}
\end{equation}

It is important to note that the generated graphs exactly adhere to the specified node count 
$N$ and magnetization $M$. However, due to the stochastic nature of edge generation, the average node degree $\kappa$ and the proportion of cross-community edges $q$ may exhibit slight fluctuations around their input values.

This generative process is similar to an Erdős–Rényi model~\cite{Erds2022}, incorporating distinct connection probabilities based on community membership. With SBM, the internal and external degrees of the communities follow a binomial distribution, with probabilities determined by the elements of $P$. 


\subsection{Estimation of $q$}\label{method:q_estim}
To derive the estimator $\hat{q}$ of $q$, we first show that the proportion of cross-community edges (equivalently, their stubs) in the network, $q$, is the harmonic mean of the proportion of stubs of cross-community edges, in the positive and negative communities, denoted as $q_+$ and $q_-$ respectively.
As such, $q_+ = E_{C}/(2E_+ + E_C)$ and $q_-=E_{C}/(2E_-  +E_C)$, where $E_\pm$ is the number of edges internally connecting each community. Given these definitions, it follows that
\begin{equation}\label{eq:q_hmean}
    q = \frac{E_C}{E} = \frac{E_C}{E_+ + E_- + E_C} =\frac{2}{\frac{1}{q_+}+\frac{1}{q_-}}.
\end{equation}
The estimator $\hat{q}$ is then obtained by substituting $q_+$ and $q_-$ with their respective estimator $\overline{\eta}_+$ and $\overline{\eta}_-$. To ensure the validity of this procedure, we demonstrate that $\overline{\eta}_+ = \sum_{i\colon x_i=1} \eta_i / N_+$ is an unbiased estimator of $q_+$, and the same rationale applies to $\overline{\eta}_-$ as an estimator of $q_-$.

For each individual $i$, $\eta_{i} = k^{C}_{i}/k_{i}$ is the proportion of stubs of the node $i$ part of a cross-community edge. By conditioning on $x_i=+1$, $k^{C}_{i}$ follows a hypergeometric distribution with $k_i$ extractions from a pool of size $2 E_+ + E_C$ (total stubs stemming from nodes voting $+1$) containing $E_C$ success states (cross-community stubs). Since $q_+ = E_C/(2 E_+ + E_C)$, then
\begin{equation}\label{eq:eta_q} 
    \E[\overline{\eta}_+] = \E_+[\eta_{i}] = \E_+\left[\frac{k^{C}_{i}}{k_{i}}\right] = 
    \E_+\left[\frac{q_+ k_{i}}{k_{i}}\right] = q_+,
\end{equation}
where $\E_\pm[\cdot] = \E[\cdot \vert x_i=\pm 1]$. Note that, in general, $\E[\overline{\eta}]$, that is $\E[\eta_i]$, equals $(N_+ q_+ + N_-q_-)/N$ rather than $q$, since $k^{C}_{i} / k_{i}$ is a mixture distribution of the two communities.

\subsubsection{Equation for the adjusted social circle}\label{method:m_corrected}
To improve the estimation of $M$, we derived an expression for the adjusted social circle within our theoretical framework and incorporated information from the unbiased estimator $\overline{x}$.
Specifically, by conditioning on $\overline{x}=\mu$, we obtain
\begin{equation}
   \E[\overline{s}|\overline{x}=\mu]=\frac{1+\mu}{2}\E_+[s_i]+\frac{1-\mu}{2}\E_-[s_i].
    \label{eq:s_given_x}
\end{equation}
We posited the assumption that $s_i$ corresponds exactly to the average voting intention of $i$'s neighbours, which implies that $s_i = x_i (1 - 2 \eta_i)$, leading to,
\begin{equation}
    \E_\pm[s_i] = \pm(1-2\E_\pm[\eta_i]) = \pm(1-2q_\pm),
    \label{eq:e_si}
\end{equation}
where the final equality follows from~\eqref{eq:eta_q}.
Consequently, equation~\eqref{eq:s_given_x} can be rewritten as
\begin{equation}
    \E[\overline{s}|\overline{x}=\mu]=\frac{1+\mu}{2}(1-2q_+)-\frac{1-\mu}{2}(1-2q_-)
    = \mu - (1+\mu)q_+ + (1-\mu)q_-.
    \label{eq:s_given_x2}
\end{equation}
We now aim to express $q_+$ and $q_-$ in terms of $q$ (which we can estimate) and $M$, to obtain a reformulation of~\eqref{eq:s_given_x2} that we can solve for $M$. Specifically, we derive
\begin{equation}
    q_\pm = \frac{E_C}{E_C + 2E_\pm} = \frac{1}{1 + 2E_\pm/E_C},
\end{equation}
by rewriting the definition of $q_\pm$. Given that $E_{\pm}=\kappa(1-q) N_\pm /2$, from \eqref{eq:epm}, $E_{C}=\kappa q N / 2$, from \eqref{eq:ec_pc}, it follows that
\begin{equation}\label{eq:qpm_mq}
    q_\pm = \frac{1}{1 + 2(1-q)N_\pm/(qN)}
    = \frac{1}{1 + (1-q)(1\pm M)/q}
    = \frac{q}{1 \pm (1-q)M}
\end{equation}
where the second relation holds since $N_\pm / N = (1\pm M)/2$.
Substituting $q_\pm$ of~\eqref{eq:qpm_mq} in~\eqref{eq:s_given_x2}, we obtain
\begin{equation}\label{eq:s_given_x3}
    \begin{aligned}
     \E[\overline{s}|\overline{x}=\mu]
     &= \mu - \frac{q(1+\mu)}{1 + (1-q)M} + \frac{q(1-\mu) }{1 - (1-q)M}
      = \mu - 2 q \frac{\mu-(1-q)M}{1-(1-q)^2M^2} \\
      &\simeq \mu - 2 q (\mu-(1-q)M)
\end{aligned}
\end{equation}
where the denominator simplifies under the assumption that $(1-q)^2M^2 \ll 1$, which holds for most realistic cases, given that $q<0.5$ and $|M|\ll1$.
Solving for $M$, it follows that
\begin{equation}
    M = \frac{\E[\overline{s}|\overline{x}=\mu] - \mu(1-2q)}{2q(1-q)}.
\end{equation}
To conclude, we substituted $\E[\overline{s}|\overline{x}=\mu]$, $\mu$, and $q$ with $\overline{s}$, $\overline{x}$, and $\hat{q}$, respectively. Thus the adjusted social circle estimator for $M$ is given by
\begin{equation}
    \hat{M} = \frac{\overline{s} - \overline{x} (1-2\hat{q})}{2\hat{q}(1-\hat{q})}.
    \label{eq:corr_sc}
\end{equation}





\subsection{Survey data preprocessing}
The USC and GfK polls contain answers to both the standard and the social circle questions. Additionally, respondents were asked to estimate the probability that they would vote. We excluded responses from individuals who indicated a zero probability of voting, as well as those intending to vote for third-party candidates. This filtering yielded 1,701 responses from the USC poll and 1,899 from the GfK poll. Respondents were then asked to provide their likelihood of voting for either Trump or Clinton. We discretized these responses, assigning a value of +1 if the likelihood was higher for Trump, and -1 if it was higher for Clinton. In rare cases where the probabilities were equal for both candidates, the vote was assigned uniformly at random.

\subsection{Analysis of weighted estimates}
To estimate the magnetisation $M$ and network polarisation $q$ on the survey data, we replaced simple means with weighted means using the weights provided with the data to improve the demographic representativeness of the samples.
Specifically, given weight $w_i$ associated to sampled individual $i$, we substituted $\overline{x}$ with the weighted mean $\overline{x}_w = \sum_{i\in Z} w_i x_i / \sum_{i\in Z} w_i $, and applied similar adjustments for $\overline{s}$, $\overline{\eta}_+$ and $\overline{\eta}_-$. Using these weighted averages, we calculated estimates from $\hat{q}$, $\overline{x}$, $\overline{s}$, and $\hat{M}$.

For all estimates, $95\%$ Confidence Intervals were calculated via percentile bootstrap method, while MSEs and MSDs were calculated following bootstrap's plug-in principle~\cite{efron1994introduction}.
In particular, we generated $B=10^5$ bootstrap samples by sampling with replacement the tuples $(x_i, s_i, \eta_i, w_i)_{i\in Z}$, thus preserving the original weight-sample pairing. For an estimator $\hat{\theta}$ of an unknown quantity $\theta$, the bootstrap replicates $\hat{\theta}^*_1, \ldots, \hat{\theta}^*_B$ were used to calculate the $95\%$ CI as $[\hat{\theta}^{*(\alpha)}, \hat{\theta}^{*(1-\alpha)}]$, where $\hat{\theta}^{*(\alpha)}$ is the $100\cdot\alpha$-th percentile of the bootstrap samples. The MSE was computed as $B^{-1}\sum_b(\hat{\theta}^*_b - \hat{\theta})^2$, while the MSD from a real value $c$ was computed as $B^{-1}\sum_b(\hat{\theta}^*_b - c)^2$.

\subsection{Statistical hypothesis testing}
The bootstrap method was also used to conduct two-sample location tests comparing the samples from the GfK and USC surveys. Four null hypotheses were tested, each challenging the equality of $\overline{x}$, $\overline{s}$, $\overline{\eta}_+$, and $\overline{\eta}_-$ between the surveys. Weighted averages, as defined above, were used to compute these values in each test. Only the procedure used for testing the location of $\overline{x}$ is detailed here, as the same methodology was applied to compare the other estimators.
First, the responses from the standard poll with associated weights for the GfK survey, $(x_i, w_i)_{i\in C}$, and for the USC survey, $(y_i, v_i)_{i\in D}$, were considered. Then, the weighted means $\overline{x}_w = \sum_{i\in C} w_i x_i / \sum_{i\in C} w_i $ and $\overline{y}_v = \sum_{i\in D} v_i y_i / \sum_{i\in D} v_i $ of the two surveys were compared using Welch's $t$-test statistic~\cite{welch1947generalization}, modified to account for weights:
\begin{equation}
    T = \frac{\overline{x}_w - \overline{y}_v}{\sqrt{\textrm{var}_w(x) + \textrm{var}_v(y)}},
\end{equation}
where $\textrm{var}_w(x) = \frac{n_C}{n_C - 1} \frac{\sum_{i\in C} w_i (x_i - \overline{x}_w)^2}{\sum_{i\in C} w_i}$, $n_C$ is the size of sample $C$, $\textrm{var}_v(y)$ is similarly defined for sample $D$.
Then, $B=10^5$ bootstrap versions $T_1^* \ldots, T_B^*$ of $T$ were generated. Each $T_i^*$ was calculated on data obtained by sampling with replacement $n_C$ tuples from $(x_i - \overline{x}_w + (\overline{x}_w + \overline{y}_v)/2 , w_i)_{i\in C}$ and $n_D$ tuples from $(y_i - \overline{y}_v + (\overline{x}_w + \overline{y}_v)/2 , v_i)_{i\in D}$, for which each sample was centred to satisfy the null hypothesis of equal means.
The $p$-value for the two-sided test was calculated as
\begin{equation}
    p = \textrm{min}(2 \cdot \textrm{min}(p_l, p_r), 1),
\end{equation}
where $p_l = (\sum_b \mathbf{1}_{(T_i^* \leq T)} + 1)/(B+1)$ and $p_r = (\sum_b \mathbf{1}_{(T_i^* \geq T)} + 1)(B+1)$, with $\mathbf{1}_{(A)}$ being $1$ if the condition $A$ is true, and $0$ otherwise.
Finally, the effect size was measured using $T/\sqrt{n_C + n_D}$,  an adaptation of Cohen's $d$~\cite{cohen2013statistical} to the weighted Welch's $t$-test statistic.

\section*{Code Availability}
The code used for this study is available at \url{https://github.com/giopal92/network_topology_social_circle}. 

\section*{Authors Contributions}
V. Loreto conceived the project. P. Gravino and G. Palermo created the theoretical framework. G. Palermo developed the simulation codes, obtained the survey data, and performed the measurements. G. Palermo and G. Prevedello performed the statistical analysis. All authors discussed and interpreted the results. P. Gravino, G. Prevedello and G. Palermo drafted the manuscript. All authors edited the manuscript. P. Gravino and G. Prevedello supervised the project.

\section*{Acknowledgements}
We sincerely thank Prof. Mirta Galesic for providing us with the survey data included in this study. This work has been supported by the Horizon Europe VALAWAI project (grant agreement number 101070930).

\bibliography{sn-bibliography}

\end{document}


\title[Article Title]{Supplementary Information.

Network topology effects on the social circle polls}

\author*[1,2]{\fnm{Giovanni} \sur{Palermo}}\email{giovanni.palermo@cref.it}

\author[1,2,3]{\fnm{Emanuele} \sur{Brugnoli}}
\author[1,2,3]{\fnm{Ruggiero D.} \sur{Lo Sardo}}
\author[1,2,3,5]{\fnm{Vittorio} \sur{Loreto}}
\author[2,4,‡]{\fnm{Giulio} \sur{Prevedello}}
\author[2,4,‡]{\fnm{Pietro} \sur{Gravino}}

\affil*[1]{\orgdiv{Physics Department}, \orgname{Sapienza University of Rome}, \orgaddress{\street{Piazzale Aldo Moro 2}, \city{Rome}, \postcode{00185}, \country{Italy}}}

\affil[2]{\orgdiv{Centro Ricerce Enrico Fermi}, \orgaddress{\street{Via Panisperna 89/A}, \city{Rome}, \postcode{00185}, \country{Italy}}}

\affil[3]{\orgdiv{Sony Computer Science Laboratories Rome, Joint Initiative CREF-SONY}, \orgaddress{\street{Centro Ricerche Enrico Fermi, Via Panisperna 89/A}, \city{Rome}, \postcode{00184}, \country{Italy}}}

\affil[4]{\orgdiv{Sony Computer Science Laboratories Paris}, \orgaddress{\street{6, Rue Amyot}, \city{Paris}, \postcode{75005}, \country{France}}}

\affil[5]{\orgdiv{Complexity Science Hub Vienna}, \orgaddress{\street{Metternichgasse 8}, \city{Vienna}, \postcode{1030}, \country{Austria}}}
\affil[‡]{These authors contributed equally to this study}

\maketitle

\section{Modularity}
In this section, we formulate the relation between the modularity $Q$ of a network with two communities and its parameters $M$ and $q$, respectively, the magnetisation and the fraction of cross-community edges. We recall that the modularity $Q$ is defined as the fraction of cross-community edges ($q$) minus the expected fraction if edges were distributed at random, which we now derive.

Given the $N>0$ nodes in the network, the number of nodes in each of the two communities is $\frac{1\pm M}{2}N$. Therefore, a node picked uniformly at random from the network will belong to one, or the other, community with probability $\frac{1\pm M}{2}$. If edges where distributed at random, the probability of selecting a cross-community edge (as opposed to one internal to a community), is equivalent to selecting two nodes uniformly at random from the network. The probability of this event is $2\frac{1+M}{2}\frac{1-M}{2} = \frac{1-M^2}{2}$. Therefore, the formula for modularity becomes
\begin{equation}
    Q = q - \frac{1-M^2}{2}
\end{equation}

\section{Assessment of estimators extended to $\kappa$ and $n$}
\begin{figure}[H]
    \begin{subfigure}[t]{0.32\textwidth}
        \centering
        \includegraphics[width=\linewidth, valign=T]{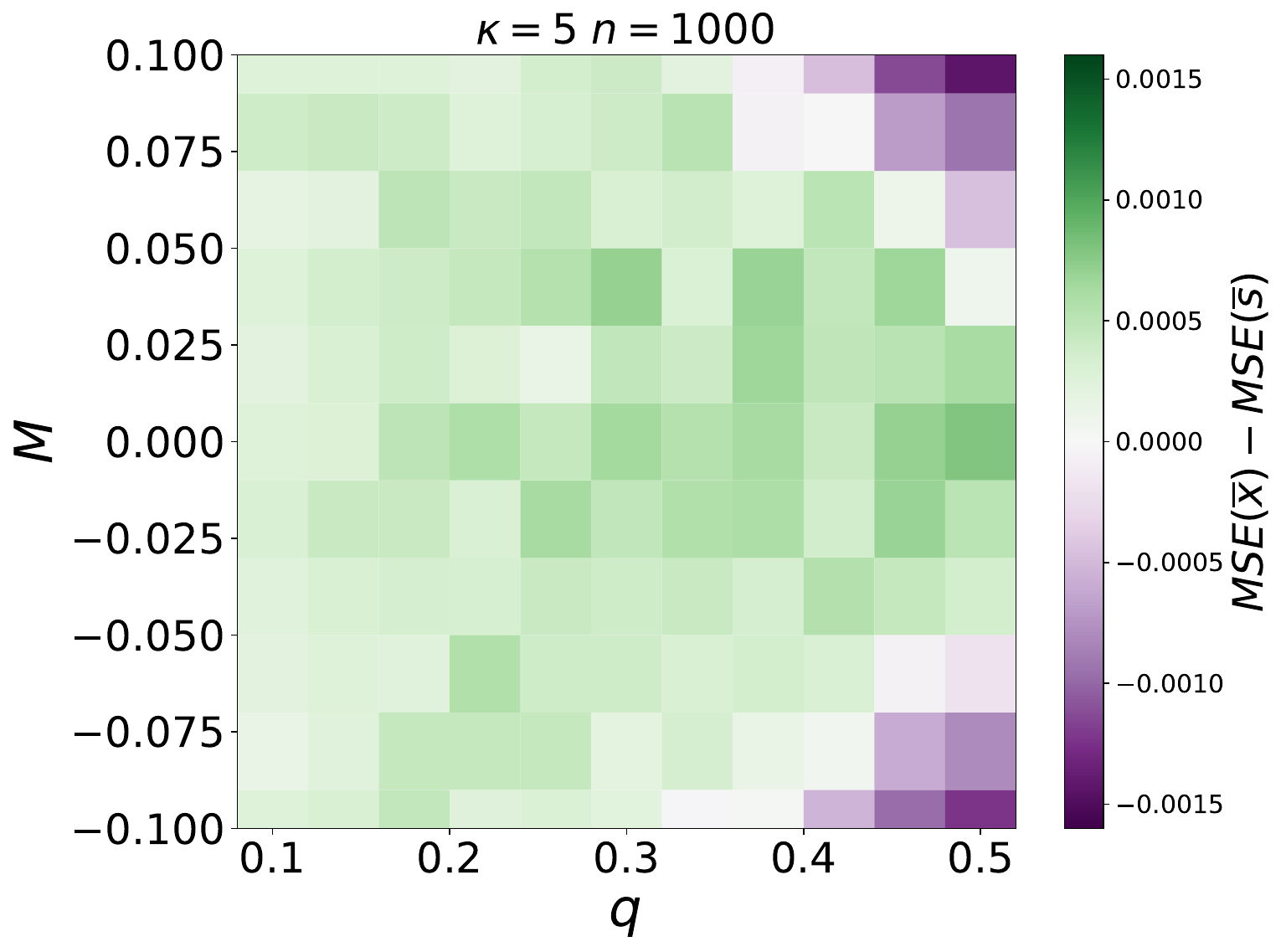}
        \subcaption{}
        \label{mse_k_5}
    \end{subfigure}
    \begin{subfigure}[t]{0.32\textwidth}
        \centering
        \includegraphics[width=\linewidth, valign=T]{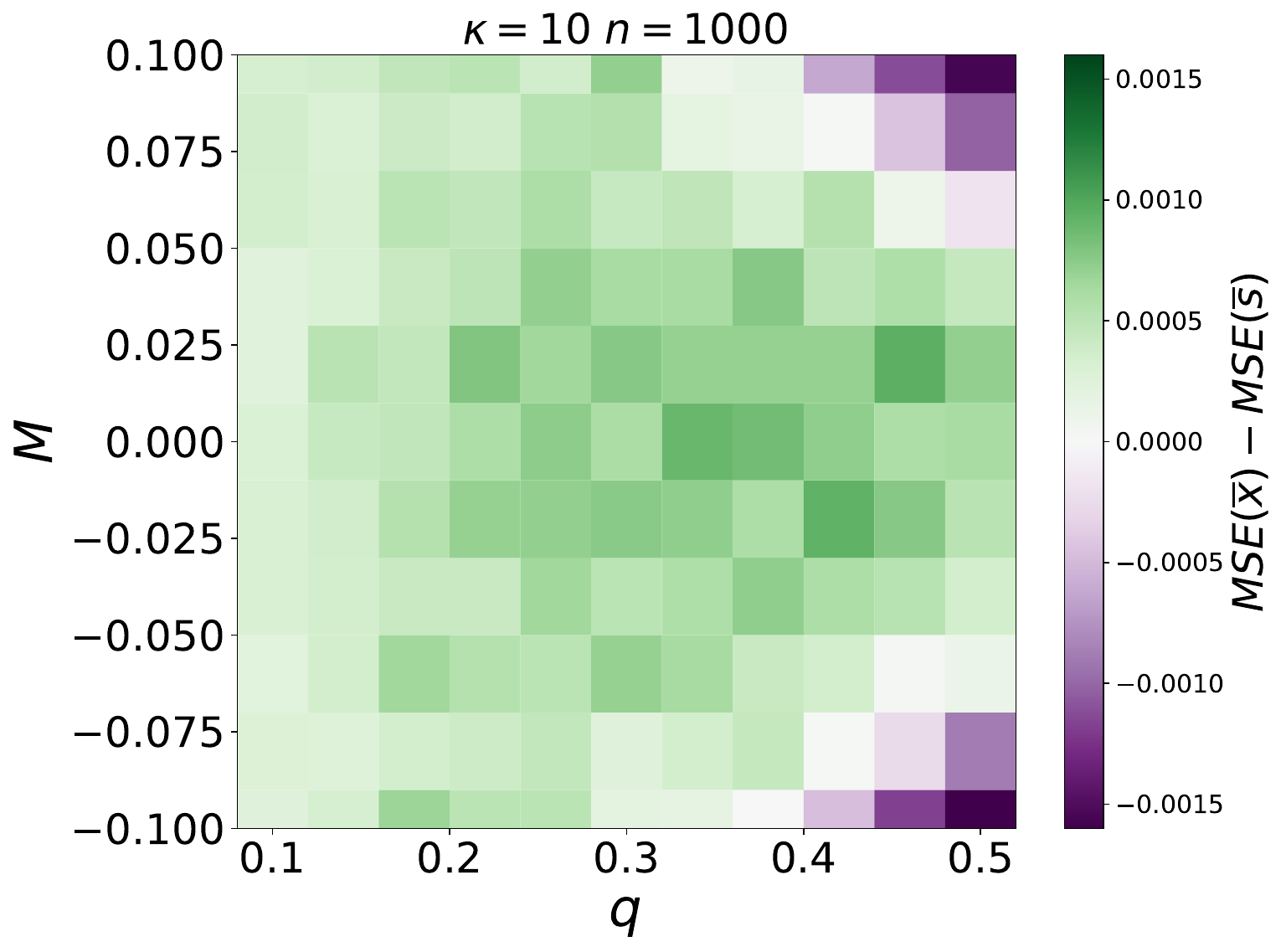}
        \subcaption{}
        \label{mse_k_10}
    \end{subfigure}
        \begin{subfigure}[t]{0.32\textwidth}
        \centering
        \includegraphics[width=\linewidth, valign=T]{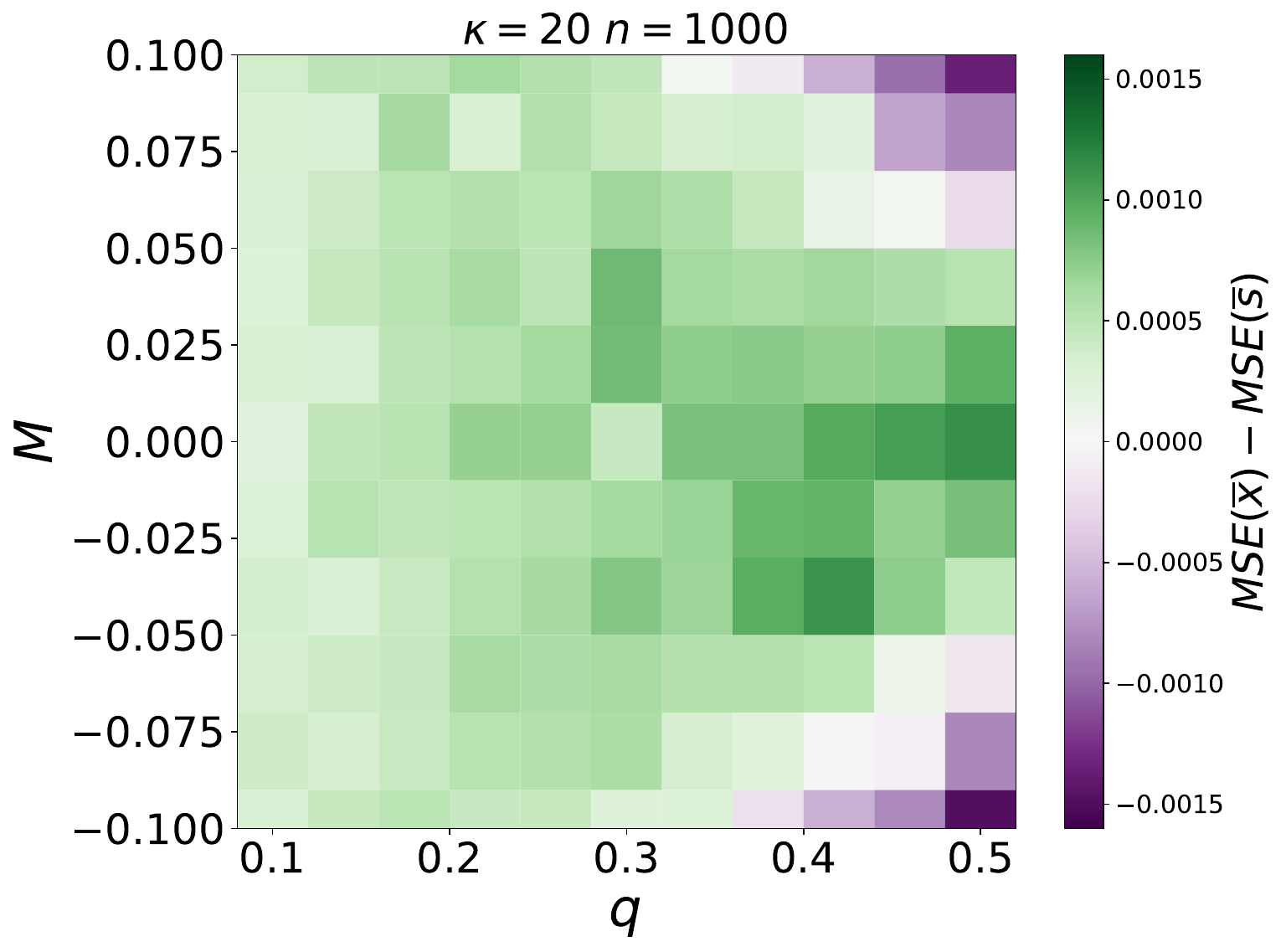}
        \subcaption{}
        \label{mse_k_20}
    \end{subfigure}

        \begin{subfigure}[t]{0.32\textwidth}
        \centering
        \includegraphics[width=\linewidth, valign=T]{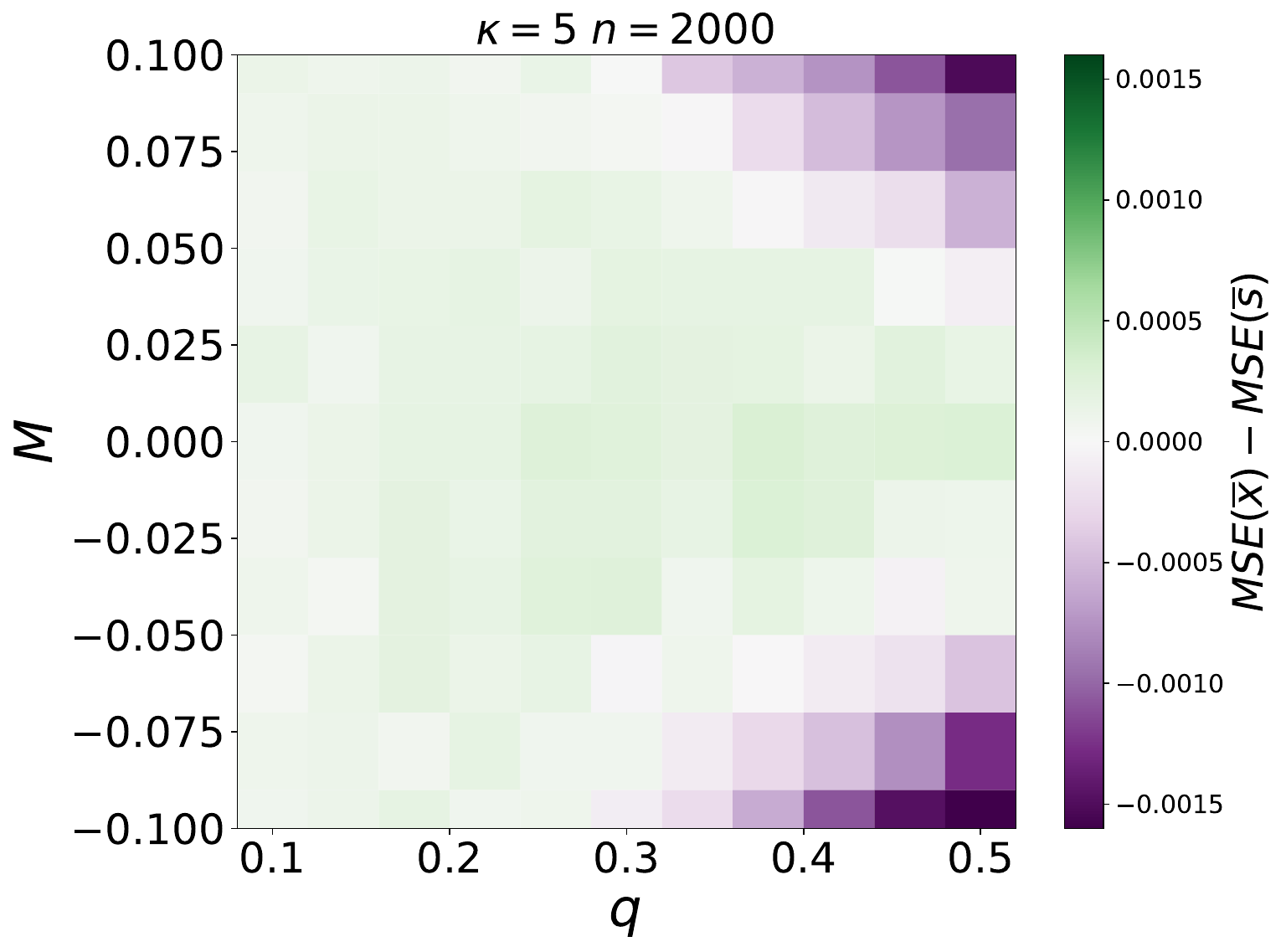}
        \subcaption{}
        \label{mse_k_5_d}
    \end{subfigure}
    \begin{subfigure}[t]{0.32\textwidth}
        \centering
        \includegraphics[width=\linewidth, valign=T]{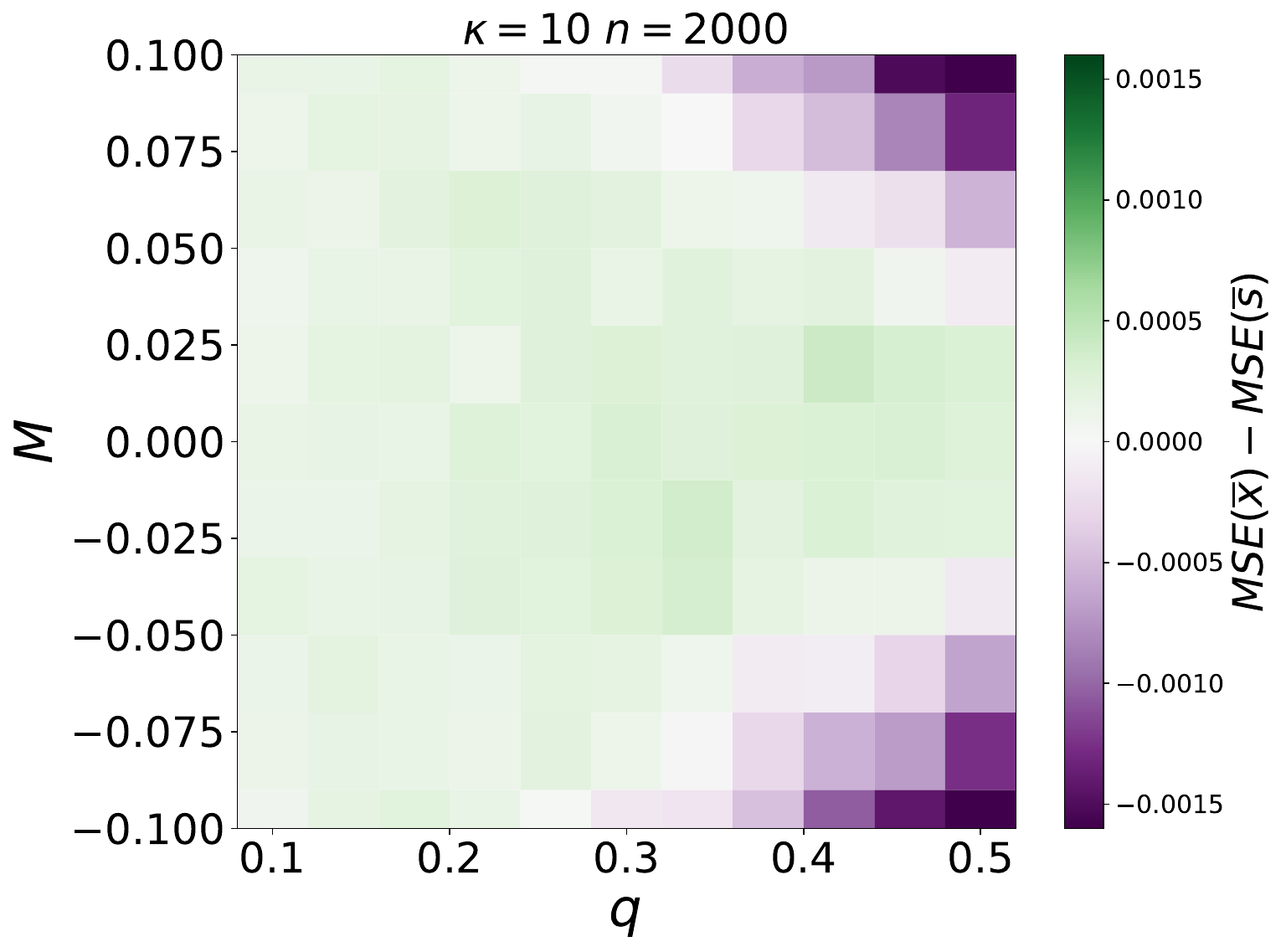}
        \subcaption{}
        \label{mse_k_10_d}
    \end{subfigure}
        \begin{subfigure}[t]{0.32\textwidth}
        \centering
        \includegraphics[width=\linewidth, valign=T]{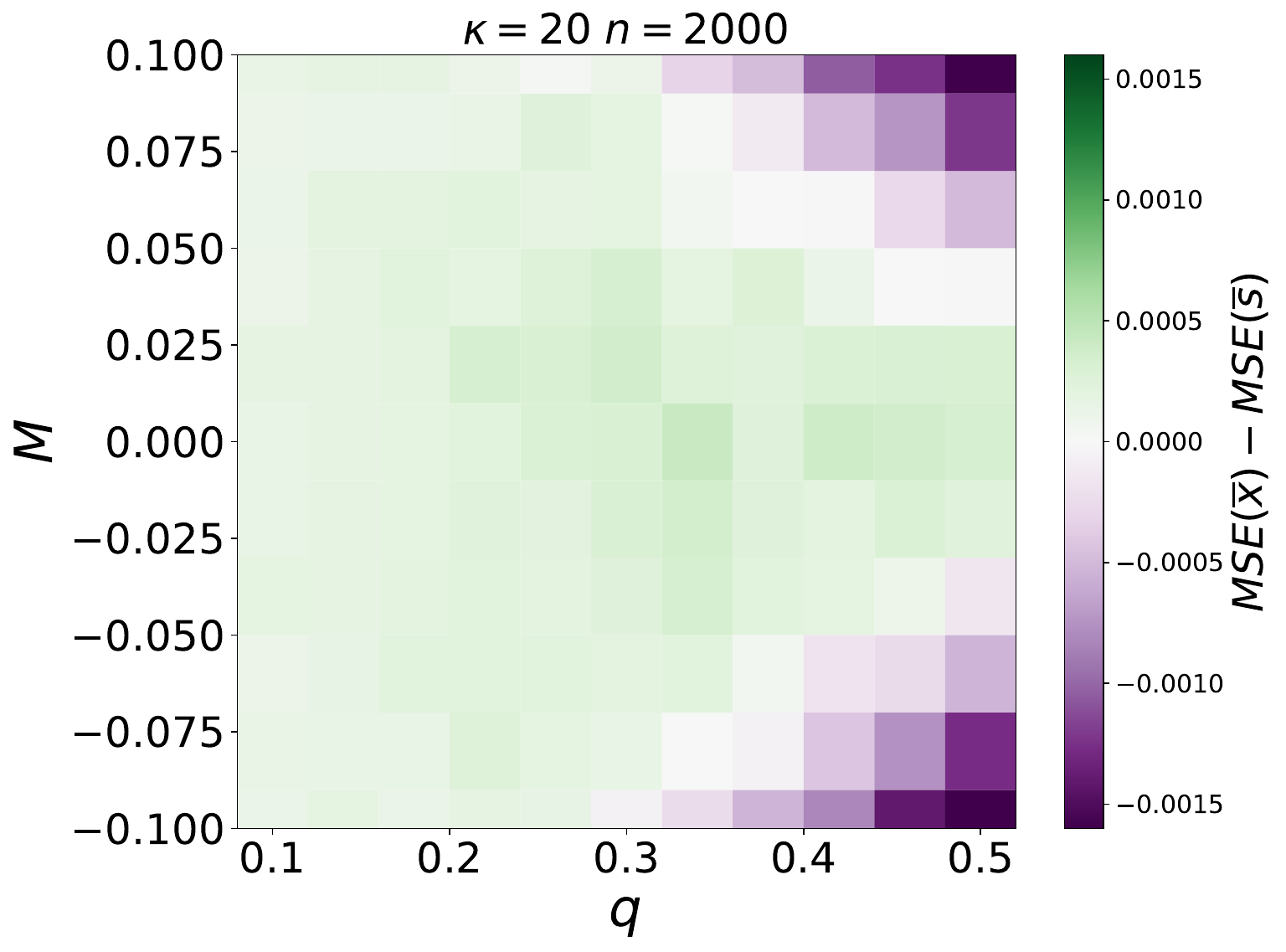}
        \subcaption{}
        \label{mse_k_20_d}
    \end{subfigure}

    \caption{Difference between MSE of the normal poll $\overline{x}$ and the social circle $\overline{s}$ as a function of $M$ and $q$ on a network with $N=10^4$. The average degree $\kappa$ spans $(5,10,20)$, while the sample size $n$ is double in the second row of the figure. Increasing $\kappa$ slightly improves the social circle performance. Increasing the sample size is highly beneficial to the normal poll, as observed in the second row ($n=2000$), where the purple area expands, and the green areas become more faded compared to the first row ($n=1000$).}
    \label{fig:heatmap1}
\end{figure}

\begin{figure}[H]
    \begin{subfigure}[t]{0.32\textwidth}
        \centering
        \includegraphics[width=\linewidth, valign=T]{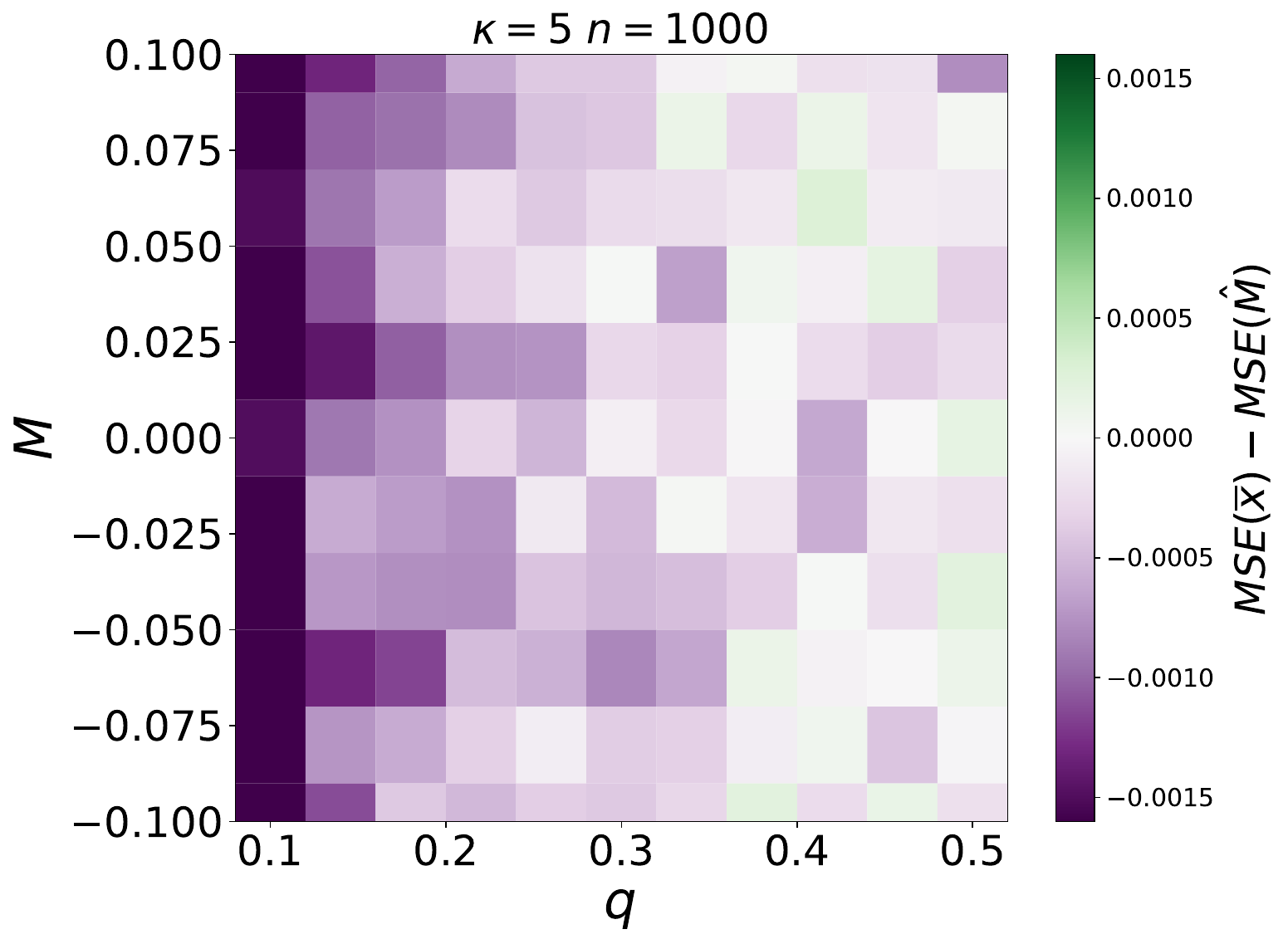}
        \subcaption{}
        \label{mse_k_5}
    \end{subfigure}
    \begin{subfigure}[t]{0.32\textwidth}
        \centering
        \includegraphics[width=\linewidth, valign=T]{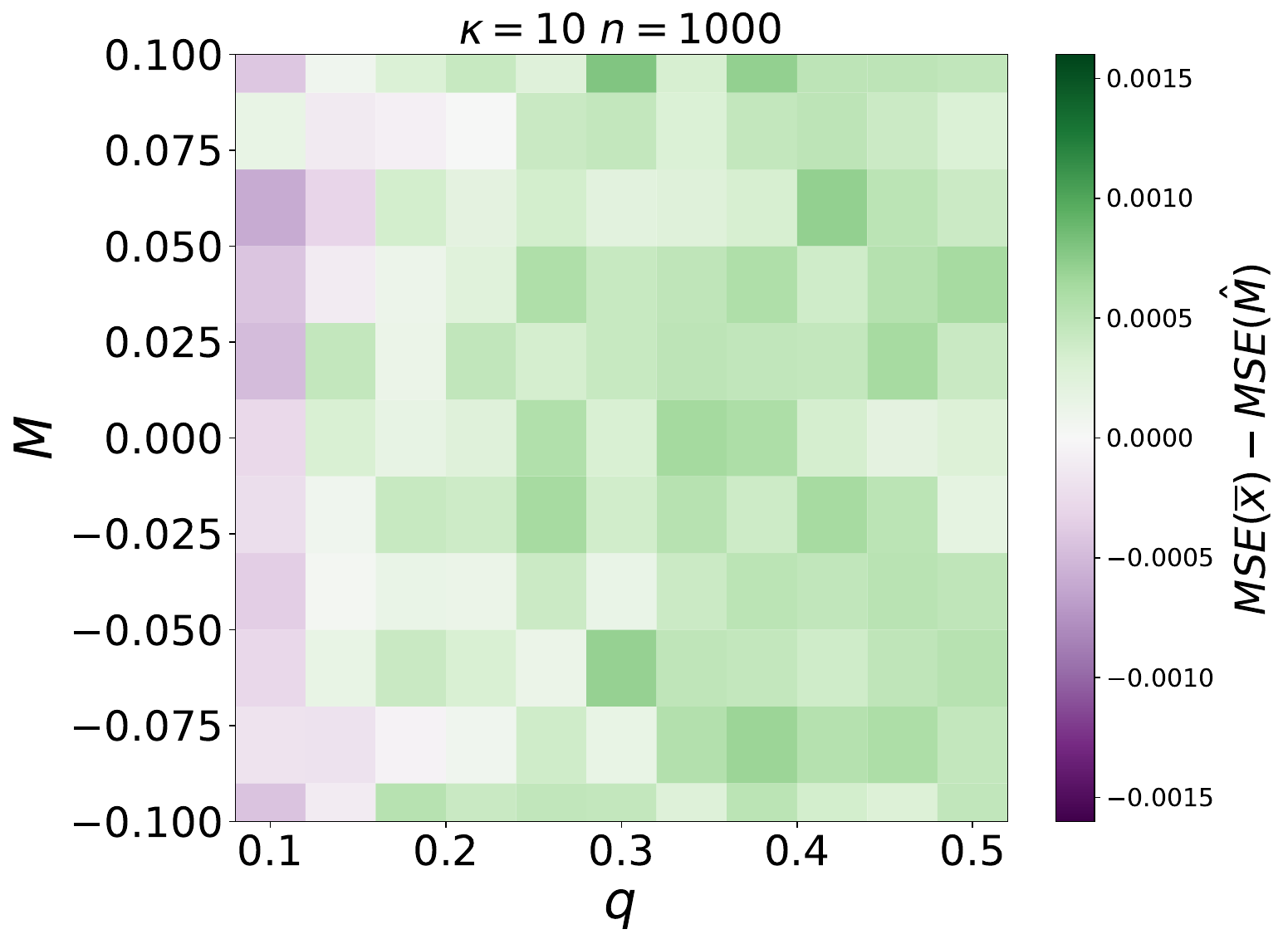}
        \subcaption{}
        \label{mse_k_10}
    \end{subfigure}
        \begin{subfigure}[t]{0.32\textwidth}
        \centering
        \includegraphics[width=\linewidth, valign=T]{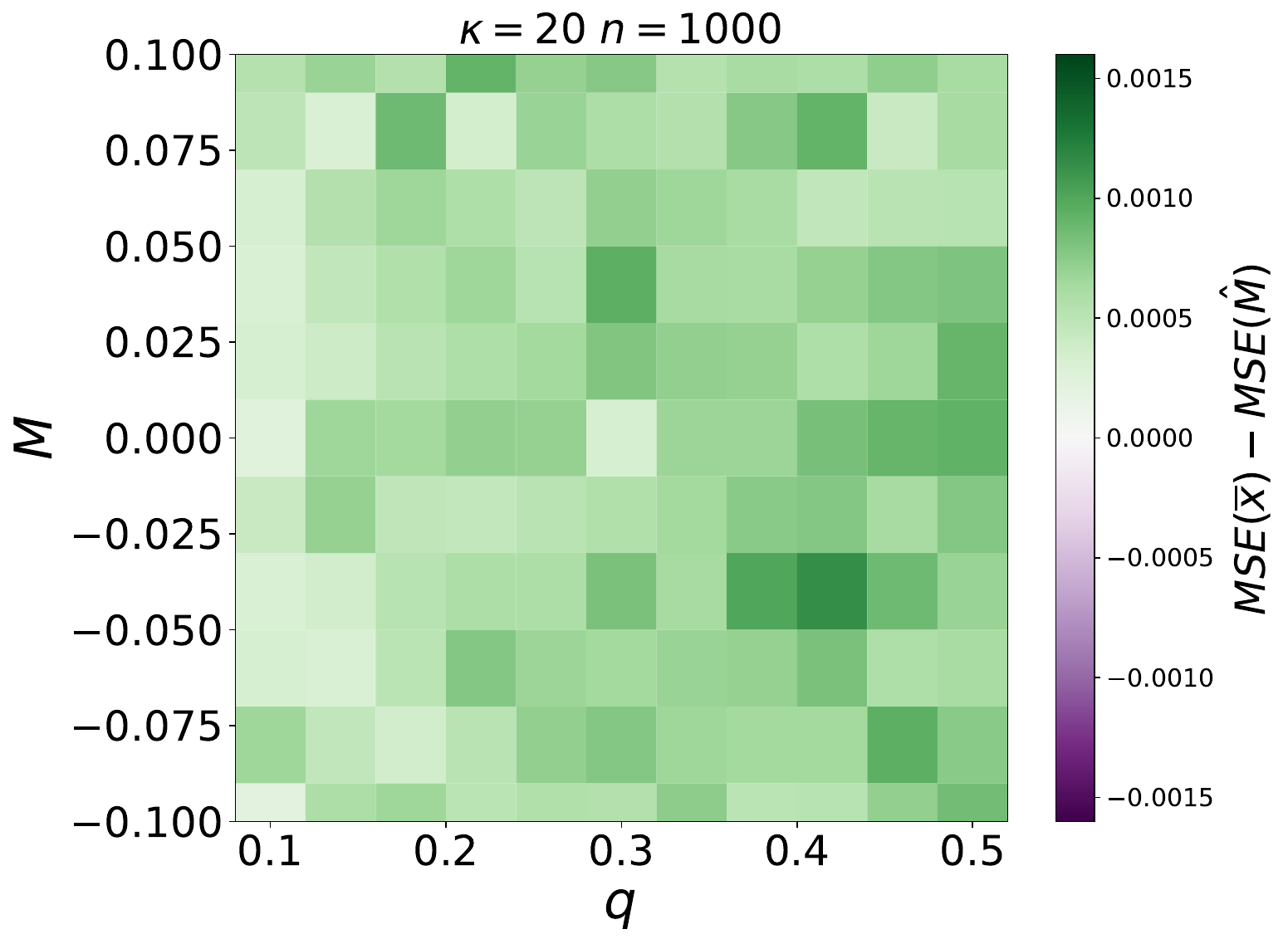}
        \subcaption{}
        \label{mse_k_20}
    \end{subfigure}

        \begin{subfigure}[t]{0.32\textwidth}
        \centering
        \includegraphics[width=\linewidth, valign=T]{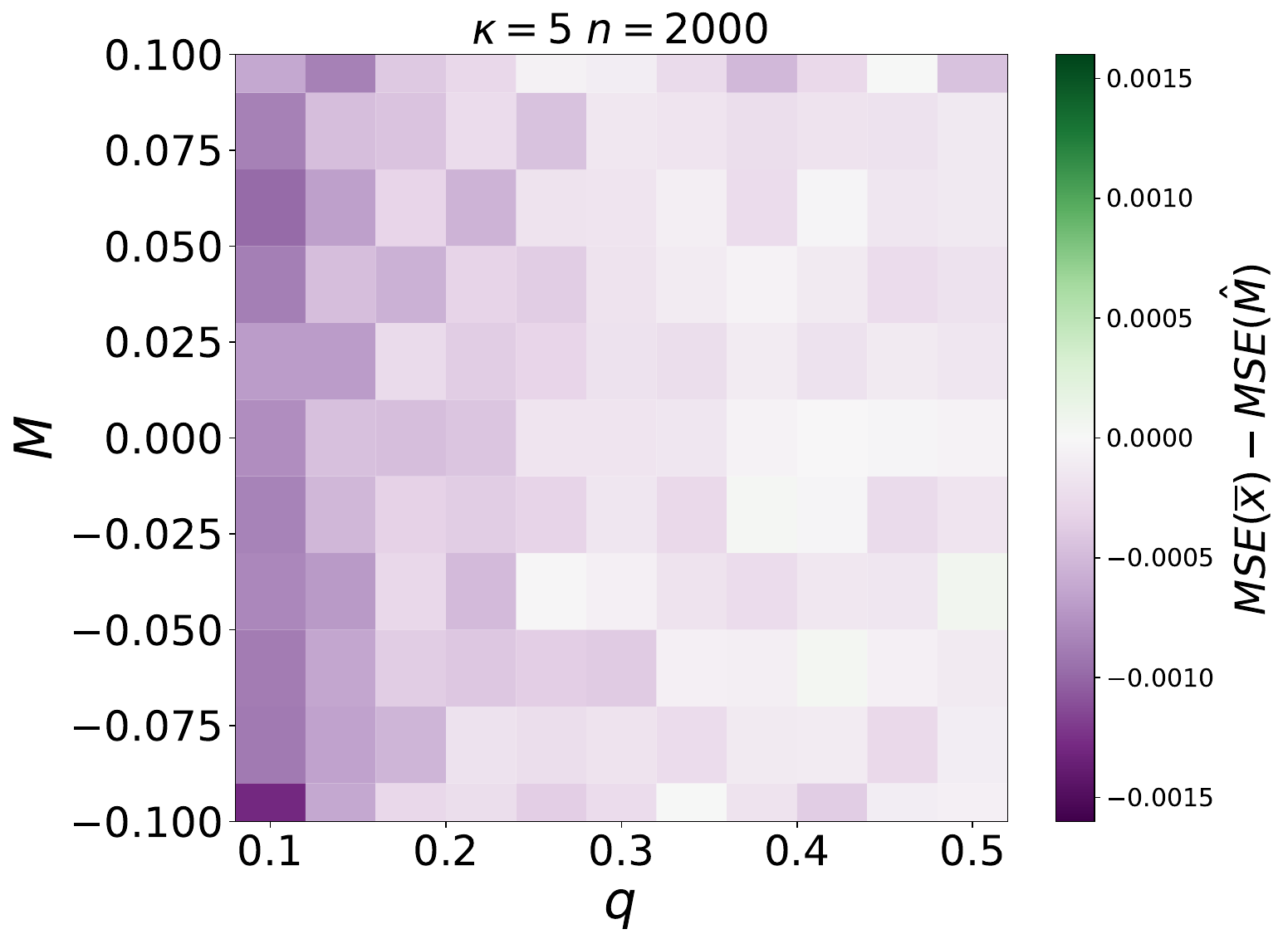}
        \subcaption{}
        \label{mse_k_5_d}
    \end{subfigure}
    \begin{subfigure}[t]{0.32\textwidth}
        \centering
        \includegraphics[width=\linewidth, valign=T]{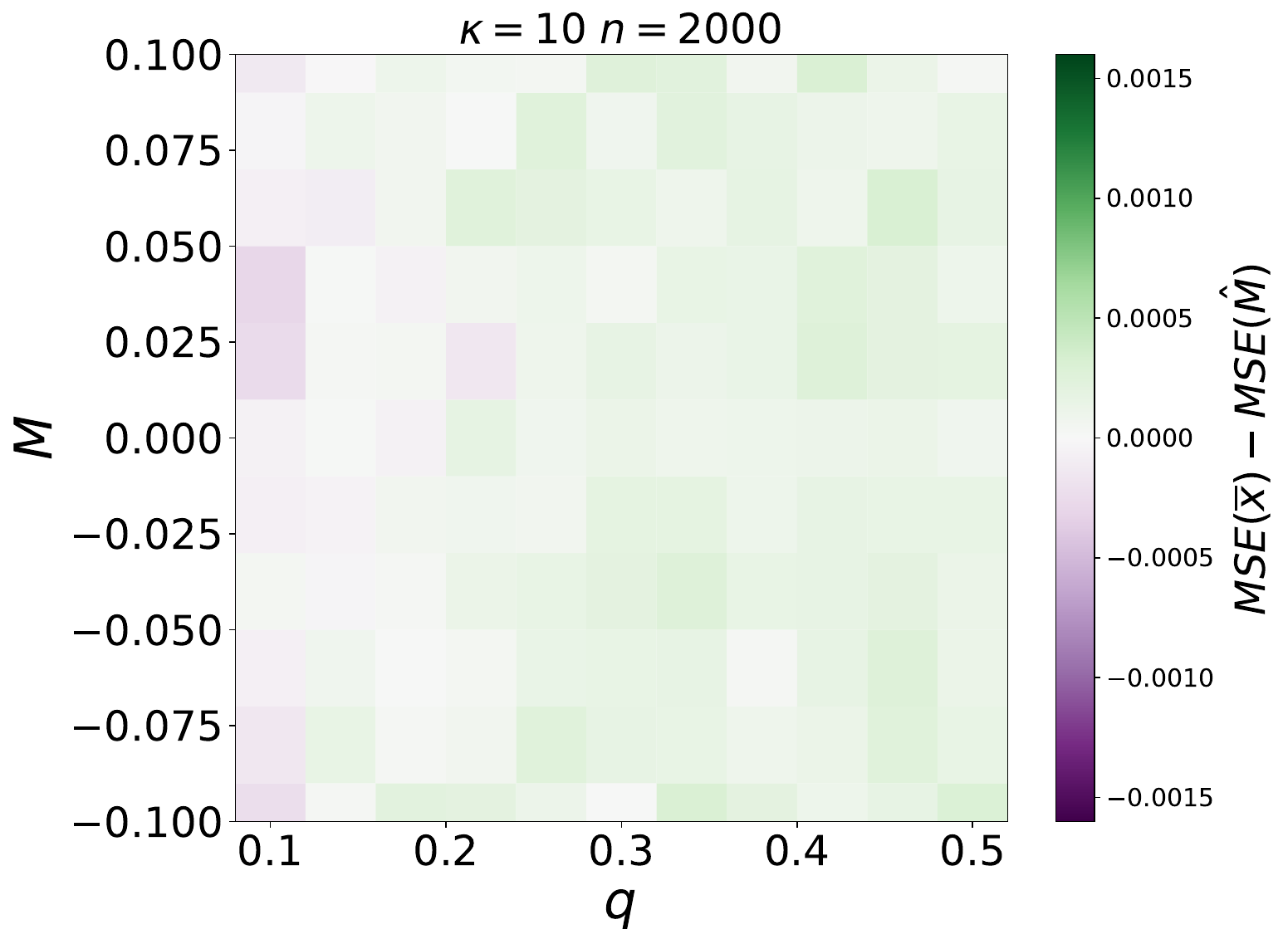}
        \subcaption{}
        \label{mse_k_10_d}
    \end{subfigure}
        \begin{subfigure}[t]{0.32\textwidth}
        \centering
        \includegraphics[width=\linewidth, valign=T]{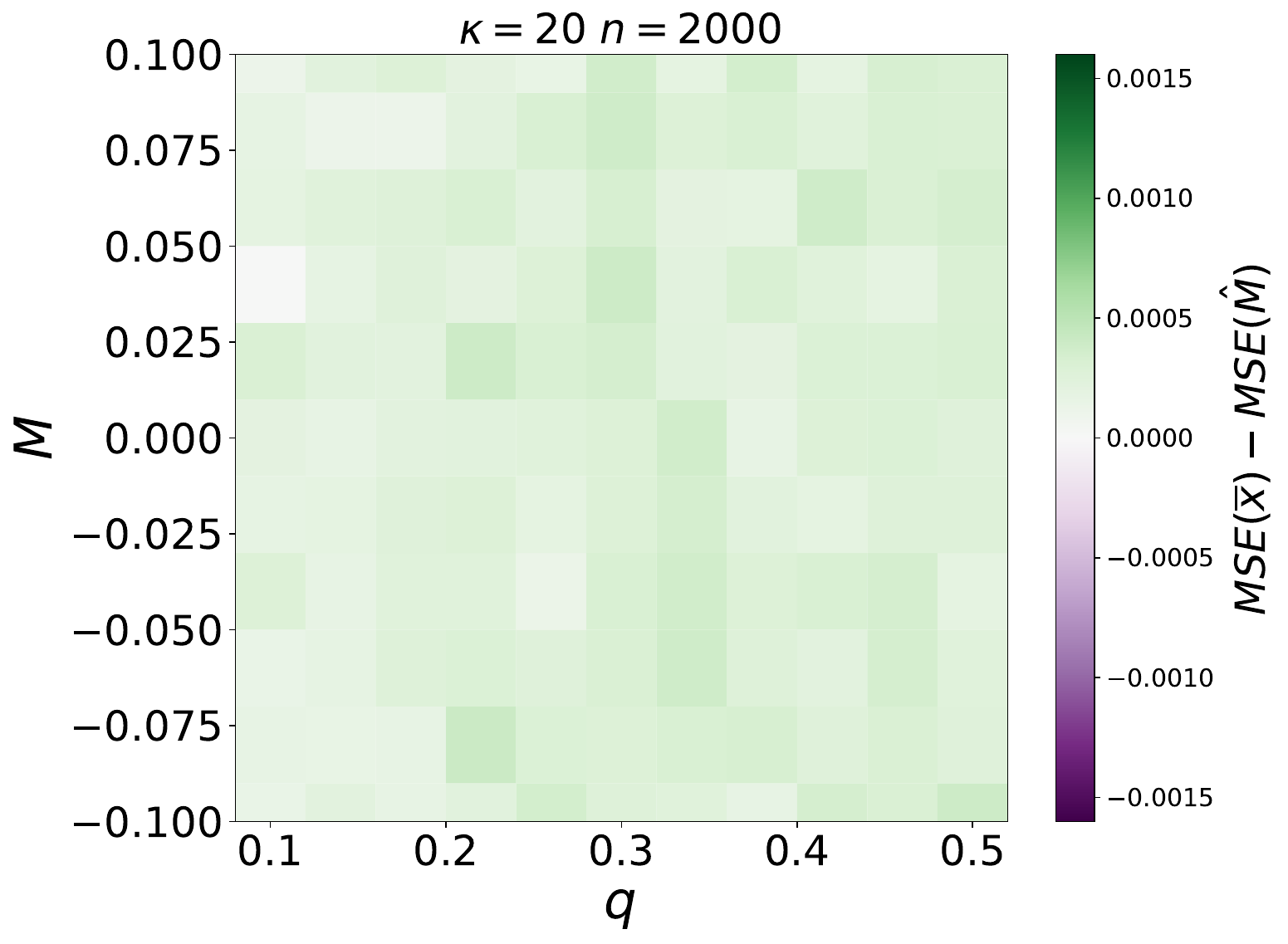}
        \subcaption{}
        \label{mse_k_20_d}
    \end{subfigure}

    \caption{Difference between MSE of the normal poll $\overline{x}$ and adjusted social circle $\hat{M}$, for the same parameters configuration used for Figure \ref{fig:heatmap1}. $\hat{M}$ behaves similarly to $\overline{s}$ in terms of dependency on the average degree of the network. It results in a poor estimator when $\kappa$ is low and outperforms the normal poll already for $\kappa \geq 10$. Again, the normal poll benefits more from a larger sample size, as in the performance analysis of $\overline{s}$.}
    \label{fig:heatmap2}
\end{figure}

In this paragraph, we assess the performance of social circle and adjusted social circle estimators in terms of Mean Squared Error (MES), compared to that of standard polling, under different values of $\kappa$, the average node degree, and $n$, the number of sampled nodes (Fig.~\ref{fig:heatmap1} and~\ref{fig:heatmap2}). While the standard estimator is not influenced by $\kappa$, the social circle improves its performance as the average number of social contacts of an individual increases. In contrast, the normal poll benefits more from a larger sample size $n$.

\section{Further statistical properties}
In this section, we compute some statistical properties (such as moments and variances) for the main stochastic variables in our study, where the network is assumed to be generated following the SBM.

\subsubsection*{Standard polling}
The variable $x_i$, for the answer to the standard poll of an individual in the network, takes values $\pm1$ with probability $\frac{1 \pm M}{2}$. Therefore,
\begin{equation}
\E[\overline{x}]=\E[x_i]=M, 
\end{equation}

\begin{equation}
    Var[\overline{x}]=\frac{1}{n}Var[x_i]=\frac{\E[x_i^2]-M^2}{n} = \frac{1-M^2}{n}.
\end{equation}
Since $\overline{x}$ is unbiased, then
\begin{equation}
    MSE[\overline{x}]
    = Var[\overline{x}]
    = \frac{1}{n}(1-M^2),
\end{equation}
and the MSE of $\overline{x}$ only depends on the magnetisation $M$ and on the sample size $n$.

\subsubsection*{Social circle polling and heterophily}
In the main text, the expressions for the expected values were already derived for $\eta_i$ and $s_i$ conditioning on community, namely $\E_\pm[\eta_i] = q_\pm$ and $\E_\pm[s_i] = \pm(1-2q_\pm)$

Following the same rationale, we derive the second moments for the conditional probabilities

\begin{equation}\label{eq:kc/k_2pm}
    \E_\pm[\eta_i^2]
    = \E_\pm[\left( \frac{k^C_i}{k_i} \right)^2]
    = q_\pm(1-q_\pm)\E_\pm[1/k_i] + q_\pm^2.
\end{equation}


The second moments lead to the variance of $\eta_\pm$

\begin{equation}
    Var_\pm[\eta_i]=q_\pm(1-q_\pm)\E_\pm[1/k_i]
    \label{eeq_var_omega}
\end{equation}

The variance of $s_\pm$ follows from the relation
\begin{equation}
    Var_\pm[s_i] = Var_\pm[\pm(2\eta_i-1)] = 4Var_\pm[\eta_i],
\end{equation}
from which, in turn, it is possible to derive the second moment $\E_\pm[s_i^2] = Var_\pm[s_i] + \E_\pm[s_i]^2$.

By the law of total expectation on the event of $i$ belonging to one group or the other, the first moment of $\eta_i$ and $s_i$ follow from the conditional expectations defined above. Thus,
\begin{equation}
    \E[\eta_i]=\frac{1-M}{2} \E_-[\eta_i] + \frac{1+M}{2} \E_+[\eta_i] = \frac{1-M}{2} q_- + \frac{1+M}{2} q_+,
\end{equation}
and
\begin{equation}
\begin{aligned}
    \E[s_i] &= \frac{1-M}{2} \E_-[s_i] + \frac{1+M}{2} \E_+[s_i]
    = -\frac{1-M}{2} (1-2q_-) + \frac{1+M}{2} (1 - 2q_+)\\
    &= M + (1-M)q_- - (1+M)q_+.
\end{aligned}
\end{equation}
The latter relation highlights how $\overline{s}$ is unbiased if and only if $(1-M)q_- - (1+M)q_+ = 0$, that is when the ratio of votes between communities (i.e., $(1-M)/(1+M)$) is the inverse of the ratio of their cross-community connections ratios (i.e., $(q_-/q_+)^{-1}$).

\section{Real survey data}
In the main text, we reported the results of experiments on real data. We used two surveys that asked both the normal and the social circle question in the occasion of the 2016 US Presidential election. For further details about the surveys, such as the exact phrasing of normal and social circle questions, refer to~\cite{Galesic2018}.

We report the number of respondents who declared a null probability of going to vote or a higher probability of voting for third parties (Table \ref{tab:poll_comparison}).



\begin{table}[h]
\centering
\small 
\renewcommand{\arraystretch}{1.2} 
\begin{tabular}{lcccc}
\hline
\textbf{Survey} & \textbf{Total} & \textbf{\footnotesize \begin{tabular}[c]{@{}c@{}}Null Prob.\\ of Voting\end{tabular}} & \textbf{\footnotesize \begin{tabular}[c]{@{}c@{}}More Likely to Vote\\ for Other Cand.\end{tabular}} & \textbf{\footnotesize \begin{tabular}[c]{@{}c@{}}More Likely to Vote\\ for Other Cand.\\ (Social Circle)\end{tabular}} \\ 
\hline
\textbf{USC} & 2,241 & 165 & 194 & 53 \\
\textbf{GfK} & 2,367 & 480\footnotemark[1] & 243 & 99 \\
\hline
\end{tabular}
\caption{Statistics for the USC and GfK surveys}
\label{tab:poll_comparison}
\end{table}

\subsection*{Preprocessing}
The US Presidential election was preprocessed to adapt to the binary vote system that we investigated and modelled.
In 2016, candidates other than Trump and Clinton received about $5\%$ of the total votes.

We first discard respondents who reported a zero probability of going to vote. Then, we binarised the vote intention into $+1$ for Trump and $-1$ for Clinton by taking the candidate that the respondent was more likely to vote for. We uniformly, at random, assign a vote intention to the people reporting equal probability of voting for Trump or Clinton. We then normalised the voting distribution of the social circle, considering only the two major candidates.
As a result, respondents had vote intention restricted to $\pm1$, and their social circles split only into Trump and Clinton voters.

\subsection*{Bootstrap distribution for $\hat{q}$}

\begin{figure}[H]
 \includegraphics[width=0.5\linewidth]{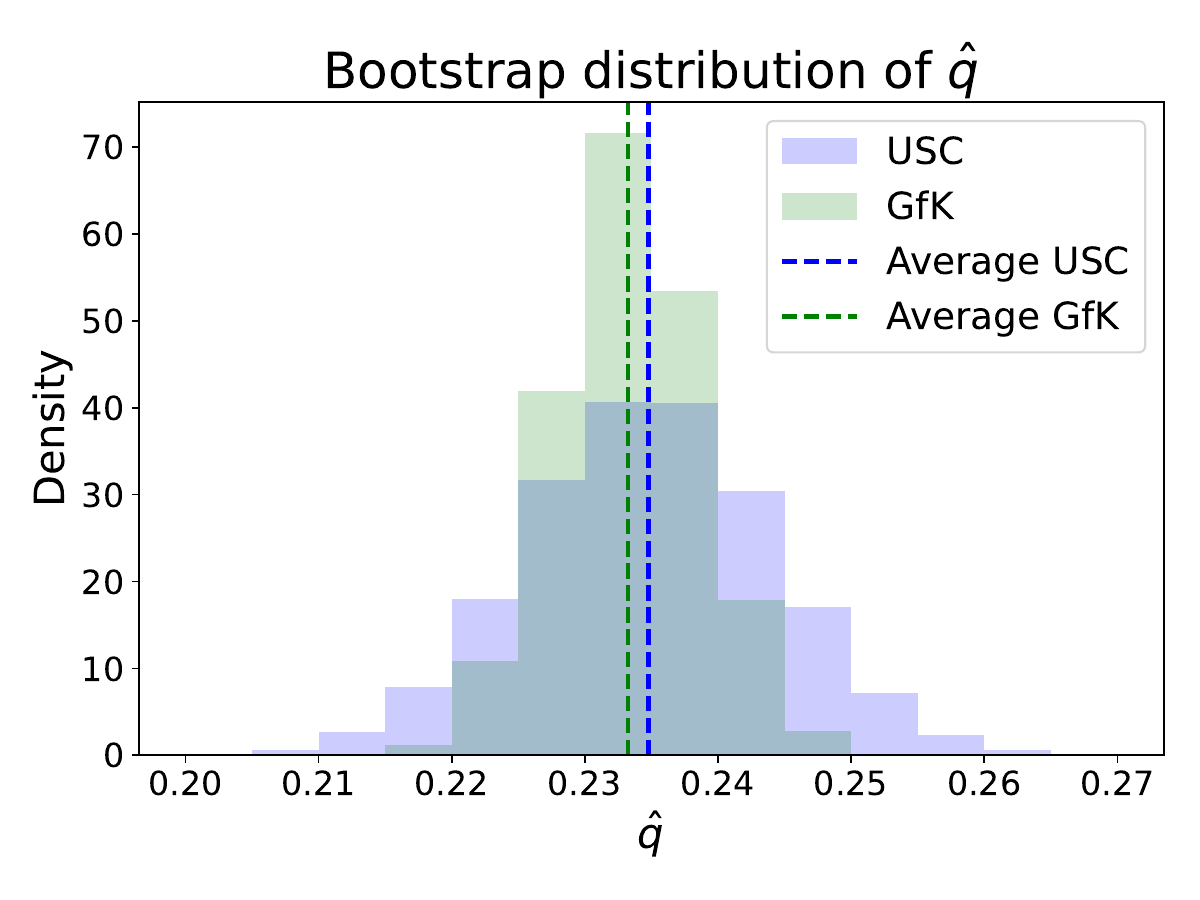}
 \centering
    \caption{Bootstrapped distributions of $\hat{q}$ for the USC and GfK surveys.}
    \label{q_real}
\end{figure}

We report the bootstrap distribution of the estimation of $q$ carried out for the USC and GfK surveys (Fig.~\ref{q_real}).

\bibliography{main/sn-bibliography}